\documentclass[iop]{emulateapj}

\shorttitle{Chandra survey of M31 GCs}
\shortauthors{Barnard et al.}

\begin{document}


\title{Chandra  identification of  26 new black hole candidates in the central region of M31}


\author{R. Barnard, and M.R.  Garcia}
\affil{Harvard-Smithsonian Center for Astrophysics (CFA), Cambridge MA 02138}
\and
\author{S. S. Murray}
\affil{Johns Hopkins University, Baltimore, Maryland; CFA}


\begin{abstract}
We have previously identified 10 M31 black hole candidates (BHCs) in M31, from their X-ray properties alone. They exhibit ``hard state''  emission spectra that are seen at luminosities $\la$10\% Eddington in X-ray binaries (XBs) containing a neutron star (NS) or black hole (BH), at luminosities that significantly exceed the NS threshold. Nine of these are associated with globular clusters (GCs); hence, these are most likely low mass X-ray binaries (LMXBs); eight  are included in this survey. We have recently discovered that analysis of the long term 0.5--4.5 keV variability of XBs via structure functions allows us to separate XBs from AGN, even though the emission spectra are often similar; this has enabled us to search for BHCs outside of GCs. We have identified 26 new BHCs (12 strong, 14 plausible) within 20$'$ of the M31 nucleus (M31*), using 152 Chandra observations spaced over $\sim$13 years; some of our classifications were enhanced with XMM-Newton observations. Of these, 7 appear within 100$''$ of M31*; this supports the theory suggesting that this region experiences enhanced XB production via dynamical processes similar to those seen in GCs.  We have found a parameter space where our black hole candidates are separated from Galactic neutron star binaries: we  show that modelling a simulated  hard state spectrum with a disk blackbody + blackbody model yields parameters that lie outside the space occupied by neutron star binaries that are modeled this way. The probability that our BHCs all lie within the NS parameter space is $\sim 3\times 10^{-29}$.
\end{abstract}


\keywords{x-rays: general --- x-rays: binaries --- black hole physics}



\section{Introduction}

We know of $\sim$20 X-ray binaries with dynamically confirmed black hole (BH) accretors; these include 15 low mass X-ray binaries (LMXBs) and 3 high mass X-ray binaries (HMXBs) in the Milky way and Magellanic Clouds \citep[see e.g.][ and references within]{remillard06}, as well as  BH + Wolf-Rayet binaries in IC10 \citep{silverman08} and NGC300 \citep{crowther10}. These systems were identified using X-ray and optical observations; the mass function is calculated from periodic radial velocity shifts in emission lines from the optical counterpart. All of the BH LMXBs identified this way are necessarily transient, because the optical spectra of bright LMXBs are dominated by the accretion disc \citep[see e.g.][]{vp94}.

We have established a method for identifying BHCs from their X-ray properties alone. This makes use of the ``low/hard'' emission state seen in BH and neutron star (NS) XBs \citep{vdk94}, that is only seen at 0.01--1000 keV  luminosities $\la$10\% Eddington in NS XBs \citep{gladstone07}; \citet{tang11} recently found that the low/hard state is limited to  luminosities $\la$10\% Eddington in BH XBs also. BH XBs may exhibit low/hard emission states at considerably higher luminosities than NS XBs, due to the higher accretor mass.  However, it is necessary to differentiate between our BHCs and distant active galactic nuclei (AGN), since AGN and XB emission spectra are often similar.

We have identified 10 BHCs from their high luminosity low states to date. Of these, 9 are associated with M31 GCs \citep{barnard08,barnard09,barnard11a,barnard2012c}, and  are therefore likely LMXBs. Only one of these is transient; however, persistently bright GC BH XBs are consistent with tidal capture of a main sequence donor \citep[][although the donor may be disrupted in the process]{kalogera04}, or with an ultra-compact system with a degenerate donor \citep{ivanova10}. We identified our first BHC outside of a GC from its X-ray spectra, long term ($\sim$12 year) behaviour, and a serendipitous HST observation \citep{barnard11b};  the faint optical counterpart ($M_{\rm B}$ $>$ $-$0.4) suggests a low mass donor for this system also. We cover 8 of the 9 GC BHCs in our survey: BHCs 1, 2, 20, 25, 28, 31, 32, and 34. The field BHC described in \citet{barnard11b} is BHC3. 
\subsection{Our GC BHCs}
We obtained accurate positions for our X-ray sources by registering 27 X-ray sources associated with M31 GCs to the M31 Field 5 B band image provided by \citet{massey06}. The r.m.s. offsets  were 0.11$"$ in R.A. and 0.09$"$ in Dec \citep{barnard2012b}; this would be extremely unlikely unless the 27 X-ray sources used for calibration were indeed associated with the GCs. 

The X-ray spectra of BH binaries are usually described with two components: a thermal component (often modelled as a multi-temperature disk blackbody), and a power law component to represent unsaturated, inverse-Compton scattering of cool photons on hot electrons \citep{remillard06}. The hard state is classified by a power law component with photon index ($\Gamma$) = 1.4--2.1, and a thermal component that contributes $<20\%$ of the 2--20 keV flux \citep{remillard06}.
 
Our very first GC BHC \citep[XB045, XBo 45 in][]{barnard08} was associated with the M31 GC B045,  named following the  Revised Bologna Catalogue v.3.4 \citep[RBC,][]{galleti04,galleti06,galleti07,galleti09}. Its $\sim$17,000 count XMM-Newton/pn spectrum was well described by an absorbed power law with line-of-sight absorption ($N_{\rm H}$) = 1.41$\pm$0.11$\times 10^{21}$ atom cm$^{-2}$, and $\Gamma$ = 1.45$\pm$0.04; 
$\chi^2$/dof = 517/487 (good fit probability, gfp, = 0.17). Adding a blackbody component improved the fit somewhat, but the thermal contribution to the flux was too small to be constrained;  we therefore considered XBo 45 to be in its hard state. Since its unabsorbed 0.3--10 keV luminosity was 2.5$\pm$0.2$\times10^{38}$ erg s$^{-1}$, 140\% Eddington for a 1.4 $M_\odot$ NS, we considered it a BHC.

The 8 remaining GCs BHCs are included in this survey, unlike XBo 45, so we don't describe each spectrum in detail here. However, we note that fitting a two component model to XB144 \citep[XBo 144 in][]{barnard09} yielded k$T$  = 0.0082$\pm$0.0016 keV, indicating a complete lack of thermal component in the spectrum. We also note that even though XB082 was best described by $\Gamma$ = 1.20$\pm$0.09, the $\chi^2$/dof for that fit was $\sim$0.9, and we were able to obtain fits where $\Gamma$=1.4 and $\chi^2$/dof $<$1 \citep{barnard11a}.

The unabsorbed 0.3--10 keV luminosities of our GC BHCs exhibiting hard state spectra range over $\sim$5--45$\times 10^{37}$ erg s$^{-1}$ \citep{barnard09,barnard11a,barnard2012c}. Comparison with the 0.5--10 keV AGN flux distribution obtained by \citet{georgakakis08} yields a 2.4$\times 10^{-36}$ probability that our GC BHCs are coincident AGN. The probability that our brightest GC BHC, XB135, is a coincident AGN is 1.2$\times 10^{-6}$. We will present evidence in a separate paper that XB135 may contain the most massive stellar mass BH known to date; it may have been formed by direct collapse of a high mass, metal poor star (R. Barnard et al., 2013,  in prep).

We found that the GCs hosting these very bright X-ray sources were significantly more massive and/or metal rich than the other GCs in M31, agreeing with previous work. However, two GCs had particularly low metalicities; one of these is B135, consistent with the direct collapse formation scenario for XB135 \citep[see][and references within]{barnard2012c}. \citet{belczynski10} have shown that while BH masses are limited to $\sim$15 $M_\odot$ for Solar metalicties, they could theoretically reach $\sim$80 $M_\odot$ for metalicities $\sim$0.01 Solar.

\subsection{Could neutron star binaries mimic the BH hard state?}

High accretion rate NS XBs exhibit multi-component emission that may appear to be hard state spectra in extragalactic X-ray sources, where the spectra have relatively few photons.
To compare the emission spectra  of our BHCs with NS XRB, we use the double thermal  (disk blackbody + blackbody) model used by Lin et al. (2007, 2009, 2012) to describe hundreds of RXTE observations of  NS XBs in all their varied  emission states. We note that the physical interpretation of their model is contradicted for persistent XBs  by a substantial body of work; however, Lin et al. (2007, 2009, 2012) have sampled spectra from the full range of NS LMXB emission states in a consistent manner, allowing us to compare our BHCs with NS XBs in a single parameter space.  

For a long time, NS XBs were divided into two types, Z-sources (high lumonisity) and atoll sources (low luminosity), based on their luminosities and color-color diagrams (CDs); the CDs of Z-sources had three branches (horizontal, normal, and flaring), and evolved along these branches without ever jumping from one branch to the other; the CDs of the atoll sources were more fragmented  \citep{hasinger89}. Furthermore, the Z-sources were split into those like Cygnus X-2, and those like Scorpius X-1 \citep{hasinger89}. It was believed that the differences were due to more than just the accretion rate, because the Z-sources varied by a factor of a few when tracing their Z-shaped CDs, while atoll source intensities  varied by 1--2 orders of magnitude \citep{muno02}. However, we now know of two transient NS systems that exhibited both types of Z-source behavior before evolving to atoll source behavior during decay \citep{homan07,chakraborty11}. 

\citet{lin07} examined the spectral evolution of two Galactic X-ray transients, Aql X-1 and 4U\thinspace 1608$-$52, over many RXTE observations covering $>$20 outbursts. They devised a new double-thermal model (disk blackbody + blackbody) to describe a NS transient soft state that is analagous to the BH soft state described by \citet{remillard06}. They have since applied their model to RXTE observations of XTE\thinspace J1701$-$462, one of the transients that exhibits Cyg-like and Sco-like Z-source behavior as well as atoll behavior \citep{lin09}, and also  to the Sco-like Z-source GC 17+2 \citep{lin12}. They have  applied their model to hundreds of RXTE spectra from Galactic NS binaries including the full gamut of NS spectral behavior. 

They  found that their disk blackbody + blackbody was unsuccessful in two situations.  Firstly, they found the hard state spectra to be power law dominated, as expected \citep{lin07,lin09}. We therefore expect our BHCs to inhabit a separate parameter space to the NS XBs because fitting the double thermal model to hard state spectra will yield unphysical results. Secondly, they found that Z-sources required a three-component spectrum (disk blackbody + blackbody + power law) on the horizontal branch \citep{lin09, lin12}. 

 \citet{lin10} also performed broadband  analysis of Suzaku and BeppoSAX observations of the persistently bright Galactic NS XB 4U\thinspace 1705$-$44, with energy ranges 0.1--600 keV and 0.1--300 keV respectively. When using a disk blackbody + blackbody model to the soft state spectra, they found  temperatures that were similar to those they observed in the RXTE observations \citep{lin07,lin09,lin12}. Fitting an additional Comptonization component, via either a power law or {\sc simpl} convolution model \citep[following][]{steiner09} resulted only in small changes in the thermal components; this is because the Comptonized component only contributed $\la$10\% of the total flux in the soft state \citep[see Fig. 6 in ][]{lin10}. Hence the parametric differences between our BHCs and the NS XBs should not be  due to differences in  energy bands used in the observations, or the lack of a third (power law) component.

Evidence against the double thermal model indicates an extended corona that contributes a substantial portion of the X-ray flux. This evidence includes the ingress times of periodic absorption dips in the X-ray lightcurves of the high inclination XBs \citep[the dipping sources, see e.g.][ and references within]{church01, church04}, and also  broadened emission lines in a Chandra grating spectroscopy of Cyg X-2 \citep{schulz09}. Furthermore, compact corona models where the inner disk temperature is tied to the seed photon energy for Comptonization are rejected for ULXs in NGC253, and the confirmed BH+Wolf-Rayet binary IC10 X-1 \citep{barnard10}, as well as  BHC3 in the steep power law state \citep{barnard11b}.

\subsection{Structure function analysis of  GC XBs}
In \citet{barnard2012c}, we applied structure function  analysis to XBs for the first time, following \citet{vagnetti11}, who created an ensemble structure function for AGN.  They used a structure function (SF) to estimate the mean intensity deviation for data separated by time $\tau$:
\begin{equation}
SF\left( \tau \right) \equiv \sqrt{\frac{\pi}{2}\left<|\log f_{\rm X} \left( t+\tau \right) - \log f_{\rm X} \left(t\right)| \right>^2 - \sigma_{\rm n}^2},
\end{equation}
where $\sigma_{\rm n}$ is the photon noise and $f_{\rm X}$ is the X-ray flux. They grouped  the SF into logarithmic bins with width 0.5; each bin in the range log($\tau$) = 0.0--3.0 contained more than 100 measurements.

Our sample consisted of 37 X-ray sources associated with objects in the RBC; these were classified by \citet{caldwell09} as 30 confirmed GCs, 4 GC candidates, 1 star, and 2 AGN. The SFs of  GC XBs with 0.3--10 keV luminosities $\sim$2--50$\times 10^{36}$ erg s$^{-1}$ tended to show significantly more variability than AGN over a wide range of time-scales. The SFs of brighter XBs generally showed comparable or less variability than AGN, despite their high signal to noise; however, their high fluxes made them unlikely AGN. Hence, SFs provide an effective mechanism for distinguishing between XBs and AGN \citep{barnard2012c}.

\subsection{Searching for field  BH XBs }

In this work, we combine these techniques to search for BHCs in the whole region covered by  152 Chandra observations from our monitoring programme; the roll angle is unrestricted, resulting an approximately circular field of view with radius $\sim20'$. We surveyed 530 X-ray sources in this region. We also obtain spectra from archival XMM-Newton observations, to strengthen the cases for several BHCs.

The 100$''$ region surrounding M31* is particularly interesting, because \citet{voss07} found an excess of X-ray binaries over the radial distribution expected from K band light (tracing stellar mass); this excess had  the distribution expected of dynamically formed XBs \citep{fabian75}. Dynamical XB formation requires stellar densities rarely seen outside GCs, and \citet{voss07} proposed that the M31 bulge is sufficiently large and dense to form a significant number of XBs dynamically. However, since the stellar velocities in the M31 bulge are $\sim$5--10 times higher than in GCs, \citet{voss07} expect only short period binaries to survive, with most of these containing BH accretors. Evidence for BHCs in this region would therefore provide strong support for their theory.

In the next section, we discuss the observations and data analysis; this is followed by our results, then by our discussion.

\section{Observations and data analysis}

\subsection{Chandra analysis}
The central region of M31 has been observed with Chandra on a $\sim$ monthly basis for the last $\sim$13 years in order to monitor transients; we exclude periods when M31 cannot be observed due to satellite constraints ($\sim$ March--April). We have analyzed 98 ACIS observations and 54 HRC observations.
We determined the position of each source from a merged 0.3--7.0 keV ACIS image, using the {\sc iraf} tool {\sc imcentroid}. The X-ray positions were registered to the Field 5 B band image of M31 from the Local group Galaxy Survey \citep[LGS, see][]{massey06}, using 27 GC X-ray sources and  following the procedure described in \citet{barnard2012b}. The r.m.s uncertainties in registration were 0.11$"$ in R.A. and 0.09$"$ in Dec.

We obtained 0.3--7.0 keV lightcurves and spectra from circular source and background regions for each source. The background region was the same size as the source region, and at a similar off-axis angle. The extraction radius varied between sources, because  larger off-axis angles resulted in larger point spread functions. We used the CIAO v4.5 software suite, with corresponding CALDB to reduce the data, and XSPEC v12.7 to analyze the spectra.

 For ACIS observations, we obtained corresponding response matrices and ancillary response files.  We initially estimated the conversion from flux to luminosity by assuming a power law emission spectrum with photon index 1.7, with $N_{\rm H}$ = 7$\times 10^{20}$ atom cm$^{-1}$, then determining the unabsorbed 0.3--10 keV luminosity equivalent to 1 count s$^{-1}$ at the location of the source. After correcting for the exposure, vignetting and background, multiplying the source intensity by this conversion factor gave the source luminosity. Source spectra with $>$200 net counts were freely fitted. When good spectral fits were found for a particular observation of a source, the parameters of these fits replaced the default parameters when estimating the source luminosity for observations of the source with $<$200 photons.


\begin{table*}
\begin{center}
\caption{ For each BHC we provide its position, and the number of ACIS and HRC observations ($O_{\rm A}$ and $O_{\rm H}$ respectively). We then give the best fit constant 0.3--10 keV luminosity, normalized to 10$^{37}$ erg s$^{-1}$, with the corresponding $\chi^2$/dof. We finally show the number of observation pairs used to create the structure function. Uncertainties are quoted at the 1$\sigma$ level.} \label{props1}
\renewcommand{\arraystretch}{.9}
\begin{tabular}{cccccccccccc}
\tableline\tableline
BHC & Position & $O_{\rm A}$ & $O_{\rm H}$ & $L_{\rm BF}/10^{37}$ & $\chi^2$/dof & $N_{\rm SF}$  \\
\tableline 
1 & 00:42:15.786+41:01:14.24 & 18 & 48 & 21.3$\pm$0.4 & 191/65 & 2145\\
2 & 00:42:18.648+41:14:01.85 & 83 & 54 & 5.54$\pm$0.06 & 1479/136 & 9316\\
3 & 00:42:22.954+41:15:35.23 & 91 & 54 & 7.57$\pm$0.06 & 23850/144 & 10440\\
4 & 00:42:26.070+41:19:15.03 & 93 & 54 & 2.53$\pm$0.04 & 3576/146 & 10731\\
5 & 00:42:28.193+41:10:00.30 & 85 & 54 & 3.85$\pm$0.05 & 486/138 & 9591\\
6 & 00:42:28.285+41:12:22.95 & 89 & 54 & 4.73$\pm$0.05 & 6113/142 & 10153\\
7 & 00:42:31.147+41:16:21.67 & 93 & 54 & 9.32$\pm$0.08 & 1123/146 & 10585\\
8 & 00:42:34.669+40:57:14.20 & 13 & 12 & 0.64$\pm$0.08 & 696/24 & 300\\
9 & 00:42:39.585+41:16:14.30 & 70 & 21 & 0.063$\pm$0.006 & 3480/90 & 4095\\
10 & 00:42:40.654+41:13:27.32 & 93 & 53 & 1.40$\pm$0.03 & 4724/145 & 10585\\
11 & 00:42:42.177+41:16:08.23 & 74 & 34 & 0.057$\pm$0.006 & 2331/107 & 5778\\
12 & 00:42:44.831+41:11:37.89 & 92 & 54 & 3.6$\pm$0.04 & 3220/145 & 10585\\
13 & 00:42:45.122+41:16:21.68 & 96 & 54 & 2.3$\pm$0.03 & 3291/149 & 11175\\
14 & 00:42:45.946+41:10:36.53 & 54 & 27 & 0.035$\pm$0.007 & 1759/80 & 3240\\
15 & 00:42:46.969+41:16:15.58 & 95 & 54 & 3.1$\pm$0.04 & 3202/148 & 11026\\
16 & 00:42:47.176+41:16:28.41 & 92 & 54 & 0.294$\pm$0.014 & 40403/145 & 10585\\
17 & 00:42:47.893+41:15:32.87 & 95 & 54 & 2.6$\pm$0.03 & 3278/148 & 11026\\
18 & 00:42:48.529+41:15:21.12 & 95 & 54 & 11.6$\pm$0.09 & 2369/148 & 11026\\
19 & 00:42:48.546+41:25:22.12 & 60 & 51 & 1.6$\pm$0.05 & 8701/110 & 6105\\
20 & 00:42:52.030+41:31:07.87 & 11 & 54 & 49.4$\pm$0.4 & 343/64 & 2080\\
21 & 00:42:52.534+41:18:54.46 & 95 & 54 & 11.79$\pm$0.09 & 1271/148 & 11026\\
22 & 00:42:54.935+41:16:03.19 & 95 & 54 & 4.8$\pm$0.05 & 18769/148 & 11026\\
23 & 00:42:57.900+41:11:04.65 & 93 & 54 & 5.3$\pm$0.05 & 826/146 & 10731\\
24 & 00:42:59.675+41:19:19.35 & 93 & 54 & 5.4$\pm$0.05 & 887/146 & 10731\\
25 & 00:42:59.872+41:16:05.72 & 95 & 54 & 5.6$\pm$0.06 & 1074/148 & 11026\\
26 & 00:43:02.937+41:15:22.53 & 95 & 54 & 3.3$\pm$0.04 & 4838/148 & 11026\\
27 & 00:43:03.220+41:15:27.69 & 95 & 54 & 1.5$\pm$0.03 & 11960/148 & 11026\\
28 & 00:43:03.876+41:18:04.91 & 94 & 54 & 4.1$\pm$0.05 & 2920/147 & 10878\\
29 & 00:43:05.667+41:17:02.43 & 77 & 29 & 0.053$\pm$0.005 & 10972/105 & 4851\\
30 & 00:43:09.866+41:19:00.76 & 85 & 48 & 0.127$\pm$0.010 & 2675/132 & 8646\\
31 & 00:43:10.622+41:14:51.30 & 90 & 54 & 13.4$\pm$0.11 & 915/143 & 10296\\
32 & 00:43:17.593+41:27:44.87 & 17 & 0 & 0.11$\pm$0.02 & 409/16 & 136\\
33 & 00:43:34.334+41:13:23.08 & 45 & 54 & 3.8$\pm$0.06 & 1380/98 & 4851\\
34 & 00:43:37.320+41:14:43.09 & 34 & 54 & 6.7$\pm$0.09 & 209/87 & 3828\\
35 & 00:43:44.593+41:24:10.25 & 32 & 39 & 0.5$\pm$0.03 & 2296/71 & 2346\\

\tableline
\end{tabular}
\end{center}
\end{table*}

For HRC observations, we we included only PI channels 48--293,  thereby reducing the instrumental background. We used the WebPIMMS tool to find the unabsorbed luminosity equivalent to 1 count s$^{-1}$, assuming the same emission model as for the ACIS observations with $<$200 photons.  We created a 1 keV exposure map for each observation, and compared the exposure within the source region with that of an identical on-axis region, in order to estimate the estimate the necessary exposure correction. We multiplied the background subtracted  source intensity by the correction factor to get the 0.3--10 keV luminosity.

We created long term 0.3--10 keV  lightcurves for each source, using the luminosities obtained from each observation as described above; all luminosities assume a distance of 780 kpc \citep{stanek98}. We only included observations with net source counts $\ge$ 0 after background subtraction. We fitted each long term lightcurve with a line of constant intensity, in order to ascertain the source variability. 

We note that the ratio of HRC to ACIS luminosity depends strongly on the spectral model, and should be 1.0 if the model is correct. Differences between actual and assumed emission spectra during HRC and faint ACIS observations may lead to  systematic offsets between luminosities from adjacent HRC and ACIS observations.

We derived SFs from the  0.5--4.5 keV fluxes of each observation of every target by assuming a power law spectrum with the same photon index as for the HRC and faint ACIS observations; an M31 X-ray source with a 0.3--10 keV unabsorbed luminosity of 1.0$\times 10^{37}$ erg s$^{-1}$, $N_{\rm H}$ = 7$\times 10^{20}$ atom cm$^{-2}$  and  $\Gamma$ = 1.7 has a 0.5--4.5 keV flux of 0.80$\times 10^{-13}$ erg cm$^{-2}$ s$^{-1}$.  

 \citet{vagnetti11} calculated the noise component from
\begin{equation}
\sigma_n^2 = 2\left<\left(\delta \log f_{\rm X}\right)^2\right> \simeq 2\left(\log e\right)^2 \left< \left( \frac{\delta f_{\rm X}}{f_{\rm X}}\right)^2\right>,
\end{equation}
assuming that $\delta f_{\rm X}/f_{\rm X}$ = $\left(1/N_{\rm phot}\right)^{0.5}$, and $N_{\rm phot}$ is the number of photons. Our lightcurves are background subtracted, and ARF-corrected; furthermore, uncertainties in the luminosities of bright sources include uncertainties in the spectral parameters.  As a result, our uncertainties are not simply due to photon counting noise. Hence in our case,
\begin{equation}
\sigma_n^2 \simeq \left(\log e\right)^2 \left< \left[ \frac{\sigma f_{\rm X}\left(t+\tau\right)}{f_{\rm X}\left(t+\tau\right)}\right]^2 + \left[ \frac{\sigma f_{\rm X}\left(t\right)}{f_{\rm X}\left(t\right)} \right]^2\right>,
\end{equation}
where $\sigma f_{\rm X}$ is the uncertainty in X-ray flux and often significantly larger than $N_{\rm phot}^{0.5}$. In the simplest case where only 1 pair has a particular separation, then 
\begin{equation}
\sigma_n^2 \simeq \left(\log e\right)^2 \left[ \left(\frac{\sigma f_1}{f_1}\right)^2 + \left( \frac{\sigma f_2}{f_2} \right)^2\right],
\end{equation}
where $f_1$ and $f_2$ are the fluxes of the two observations.

\subsection{XMM-Newton analysis}

In addition to our Chandra observations, we also obtained spectra for some of our BHCs from archival XMM-Newton observations. Their higher sensitivity and longer exposure times often yielded superior spectra for bright sources. We used only the pn instrument, because the MOS detectors are more prone to pile up. We used SAS v10 for the data reduction.

 We filtered out  background  flares by creating a lightcurve with the selection ``(PATTERN==0)\&\&(FLAG==0)\&\&(PI in [10000:12000])'' and 100 s binning, then excluding intervals with $>$0.4 count s$^{-1}$. Source and background spectra were obtained from circular regions with 20--40$"$ radius, along with corresponding response files; the filter ``(PATTERN$<$=4)\&\&(FLAG==0)'' was applied along with the good time filter and extraction region.


\begin{figure}
\epsscale{1}
\plotone{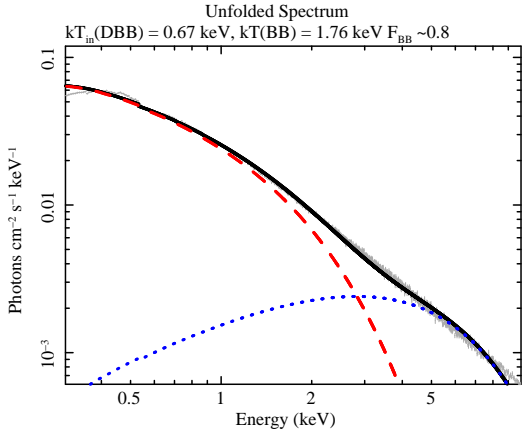}
\caption{Faked spectrum created using an absorbed power law model that represents the hard state: $N_{\rm H}$ = 7$\times 10^{20}$ atom cm$^{-2}$, $\Gamma$ = 1.7. The best fit double thermal model gives a 0.67 keV disk blackbody and a 1.76 keV blackbody; the blackbody component contributes $\sim$80\% of the 2--10 keV flux.   }\label{example}
\end{figure}


\subsection{Identifying black hole candidates}

Our BHC identification process involved three steps. First, we had to establish that the X-ray source was an XB rather than an AGN. Then we checked the ACIS observations of each XB, to see if it exhibited a high luminosity hard state ($\Gamma$ $\le$ 2.1 at a 0.3--10 keV unabsorbed luminosity $>$3$\times 10^{37}$ erg s$^{-1}$, i.e. 10\% Eddington for a 2 $M_{\odot}$ NS). Finally we found the highest quality observation that exhibited a high luminosity hard state, fitted it with the double-thermal model, and compared the results with the known range of parameters exhibited by the NS systems studied by Lin et al. (2007, 2009, 2012).

Likely XBs were identified from their associations with GCs, variable SFs or relatively high fluxes. There are $\sim$170 GCs within our field \citep{peacock11}; using the 0.5--10 keV AGN flux distribution \citep{georgakakis08}, we expect to find 0.0005 AGN with flux $>$1.4$\times 10^{-13}$ erg cm$^{-2}$ s$^{-1}$ (10$^{37}$ erg s$^{-1}$) within 1$"$ of any GC. Hence any bright X-ray source associated with a GC is a likely XB. We also identify XBs using SFs that are strikingly more variable than the ensemble AGN SF. Furthermore, only $\sim$1 AGN per square degree is expected with a flux equivalent to $\sim 5\times 10^{37}$ erg s$^{-1}$; hence brighter X-ray sources are likely XBs too.

Lin et al. (2007, 2009) found that they were unable to successfully fit hard state spectra with their double thermal model. This is because the disk blackbody component must account for the low-energy flux, resulting in an unphysically low temperature. To demonstrate this, we created a fake hard state spectrum  using the {\sc xspec} command {\sc fakeit}; we used  a power law with $\Gamma$ = 1.7, with absorption equivalent to 7$\times 10^{20}$ atom cm$^{-2}$. We present the best fit double thermal model for this spectrum in Fig.~\ref{example}, which consists of a 0.67 keV disk blackbody and a 1.76 keV blackbody; the blackbody component contributes $\sim$80\% of the 2--10 keV flux. By contrast, the soft states of NS XBs studied by Lin et al. (2007, 2009, 2012) yielded disk blackbody  temperatures $\ga$1 keV, with the blackbody component contributing less than 50\% of the 2--10 keV flux. 

Since Lin et al. (2007, 2009, 2012) failed to fit hard state spectra with the double thermal model, we expect disk blackbody + blackbody fits to our BHCs to inhabit a different parameter space from the NS binaries if they really are in the hard state. 
If a BHC exhibits a blackbody temperature $<$1.5 keV, a disk blackbody temperature $<$1.0 keV ($<$1.2 keV for disk blackbody 2--10 keV luminosity $>$2$\times 10^{37}$ erg s$^{-1}$), and/or a blackbody contribution to the total 2--10 keV flux $>$45\% at a 3$\sigma$ level, then we consider it a strong BHC; otherwise, we label it a plausible BHC. These criteria are drawn from the hundreds of RXTE spectra of  Aql X-1, 4U\thinspace 1608$-$52, XTE\thinspace J1701$-$462, and GX 17+2  analyzed by \citet{lin07,lin09,lin12}.


\begin{figure*}
\epsscale{1}
\plotone{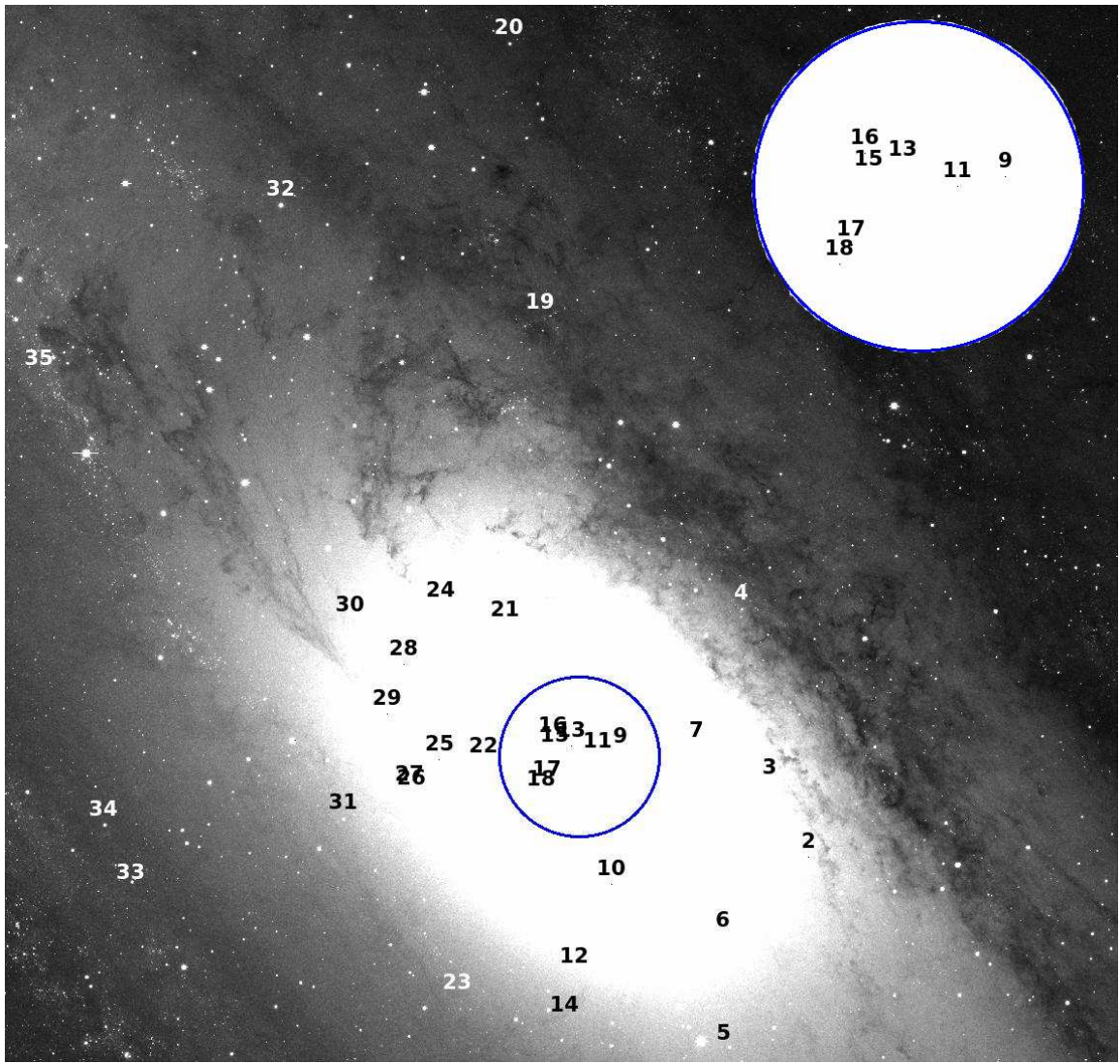}
\caption{Detail of the B band image of M31 Field 5 from the LGS \citep{massey06}, superposed with the X-ray positions of our BHCs. The blue circle has a 100$''$ radius ($\sim$380 pc), and denotes the region where Voss et al. (2007) predict enhancement for dynamically formed XBs, many of which are expected to contain black holes. North is up, and East is left. A magnified view of the central 100'' region is presented at the top right of the image. BHCs 1 and 8 are located outside the Field 5 image.  }\label{lgsbh}
\end{figure*}

\section{Results}
\label{res}

Table~\ref{props1} gives the location of each of our BHCs, followed by the number of observations with ACIS and HRC. Next we provide the best fit line of constant 0.3--10 keV luminosity  over $\sim$13 years, with the corresponding $\chi^2$ and number of degrees of freedom (dof). Finally, we show the number of observation pairs  used to create the structure function. Usually this is $N_{\rm obs}\left(N_{\rm obs}-1\right)/2$, where $N_{\rm obs}$ is the total number of observations;  however, BHCs 7, 29, 30, and 35 included observations with zero luminosity that could not be included in the SF. All uncertainties in this work are quoted at the 1$\sigma$ level unless specified otherwise.

For nearly all BHCs, closely-spaced ACIS and HRC observations give consistent luminosities, although some small systematic offsets occur due to differences between assumed and actual spectra when fitting is not possible. However, BHC32 appears to have substantial signal in the HRC observations even when this transient source is in quiescence; the HRC is considerably more sensitive than ACIS to low energy photons, hence the  HRC obsevations of BHC32 must be contaminated by soft X-rays. BHC32 is located near 2 variable stars (16$"$ and 17$"$ distant); since the extraction radius for this souce is 15$"$ due to the high off-axis angle, our HRC observations are likely contaminated by one or both foreground stars.  As a result, we excluded all HRC observations of BHC32 from our analysis.


\begin{figure*}
\epsscale{1}
\plotone{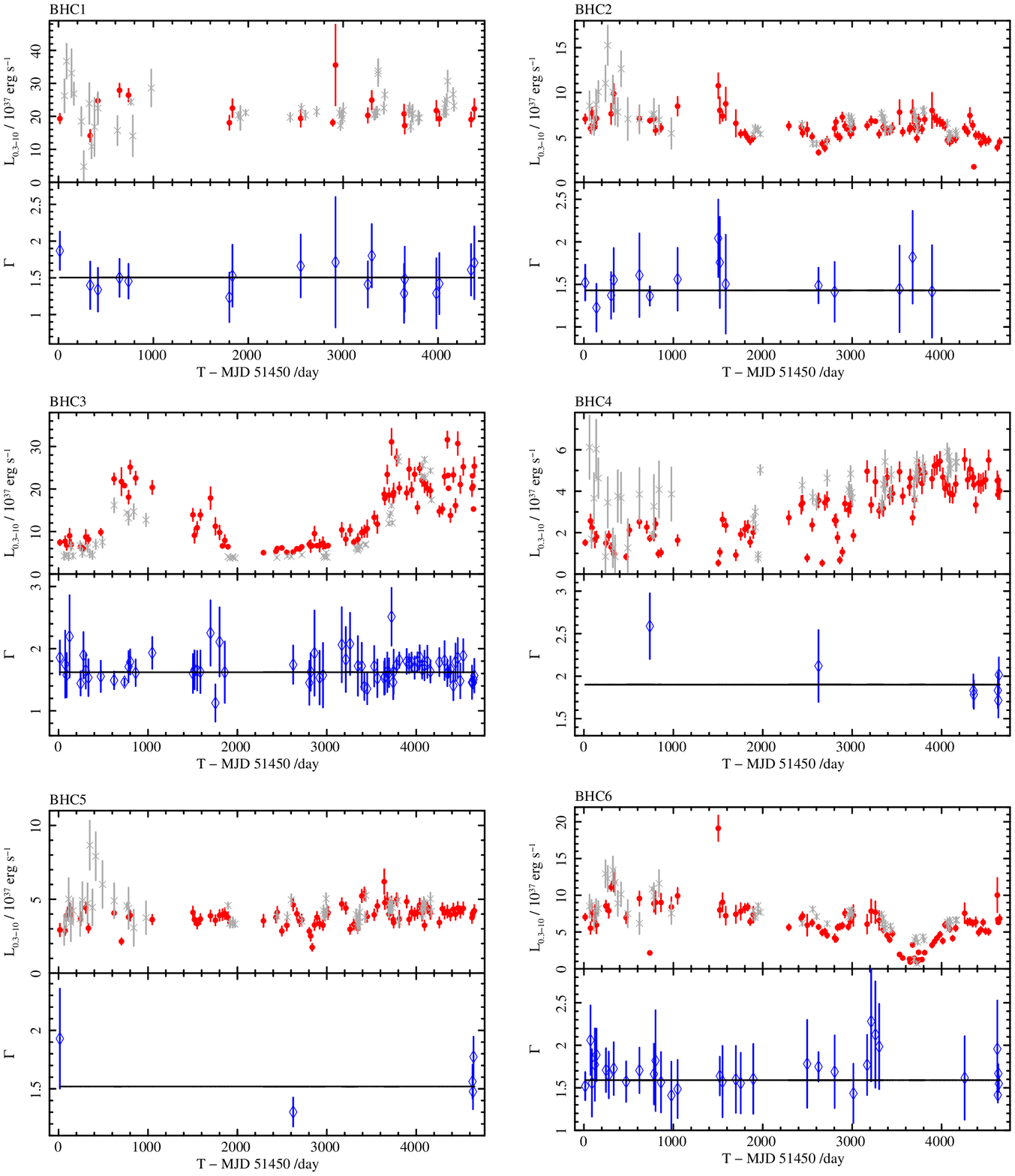}
\caption{The top half of each panel shows the 0.3--10 keV luminosity of in each Chandra observation made over $\sim$13 years; the red circles and grey crosses represent ACIS and HRC observations, respectively. The bottom panel shows $\Gamma$ for the best fits  to freely-fitted spectra with a power law emission model; the mean $\Gamma$ is indicated by a line. Uncertainties are at the 1$\sigma$ level for luminosities, and 90\% confidence level for $\Gamma$. An extended, color version of this figure is available in the electronic edition.}\label{lcs}.
\end{figure*}

\subsection{BHC locations}

Figure~\ref{lgsbh} shows a detail of the B band image of M31 Field 5 from the LGS (Massey et al., 2006), superposed with the positions of our BHCs; north is up, east is left. The circle encloses the region within 100$''$ ($\sim$380 pc) of M31*; this region is enlarged in the top right portion of the figure. BHCs 1 and 8 lie below the southern edge of Field 5. 

\subsubsection{The central 100$''$ region}
We find 7 BHCs within  100$''$ of M31$^*$, 20\% of our total sample: BHCs 9, 11, 13, 15, 16 , 17, and 18. This result appears to be consistent with the predictions of \citet{voss07}.  BHC9 is  a recurrent  transient with a peak 0.3--10 keV luminosity of $\sim$8$\times 10^{37}$ erg s$^{-1}$.  BHC11 was bright for  $\sim$1--2 years, then turned off. BHC18 appears to be persistently bright ($\sim 10^{38}$ erg s$^{-1}$), and may vary rapidly if the HRC luminosities are correct; however, the emission from BHC18  is variable ($\Gamma$ $\sim$1.4--2.4 for bright ACIS observations), meaning that the HRC luminosities may be systematically offset due to differences between assumed and true spectra.

  BHCs 13, 15, 16, and 17 are extremely variable, with BHCs 13, 15, and 17 luminosities ranging over $\sim$1--6$\times 10^{37}$ erg s$^{-1}$, and BHC16 exhibiting luminosities from $\sim$5$\times 10^{36}$ to $\sim$2$\times 10^{38}$ erg s$^{-1}$. We observed similar behavior in the X-ray source associated with B158 \citep{barnard2012c}, a high inclination LMXB with a $\sim$10000 s period \citep{trud02}, and an asymmetric, precessing disc \citep{barnard06}. Such behavior is seen in low mass ratio (short period) systems where the outer disc reaches the 3:1 resonance with the donor star, causing additional tidal torques that lead to elongation and precession of the disc \citep[ see e.g.][]{osaki89, wk91, od01}. B158 exhibited rapid variation over 4--40$\times 10^{37}$ erg s$^{-1}$, which we speculated might be due to varying accretion rates over the disc precession cycle; we suggested that other XBs exhibiting such behaviour may also be in precessing (i.e. short period) systems \citep{barnard2012c}. Therefore we propose that BHCs 13, 15, 16, and 17 are short period BH XBs, as predicted by \citet{voss07}.

\subsubsection{Further GC associations}
Two of our new BHCs are associated with confirmed old GCs; BHC4 is associated with B096 (following the RBC naming convention), and BHC24 with B143. \citet{caldwell11} rank B096 as the 36th most massive and 34th most metal rich out of 379 GCs  ([Fe/H] = $-$0.28); B143 is ranked 52nd most massive and 17th most metal rich, with [Fe/H] = $-$0.07. We have found that the GCs hosting our previous BHCs were particularly massive, or metal rich, or both (Barnard et al., 2011a, 2012); hence it is encouraging that the new GC BHCs are associated with GCs that are amongst the most massive and most metal rich  of the M31 GCs studied by Caldwell et al. (2009, 2011).   

In \citet{barnard2012c} we classified the semi-regular transient associated with HII rich GC candidate  SK059A as a possible HMXB associated with the HII, because the intervals between  its outbursts were consistent with $n$ cycles of 120 days; hence the accretor could make an eccentric, $\sim$120 day orbit around the high mass donor, with mass transfer restricted to times near periastron. This system is BHC30.


\begin{figure}
\epsscale{1}
\plotone{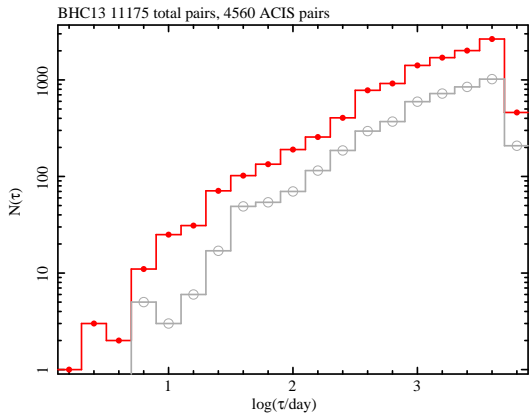}
\caption{Histogram showing the number of observations with separation $\tau$ for BHC13, which was most frequently observed. The histogram with filled circles includes ACIS and HRC observations, while the histogram with open circles uses only ACIS observations. }\label{hist}
\end{figure}
\begin{figure*}
\epsscale{1}
\plotone{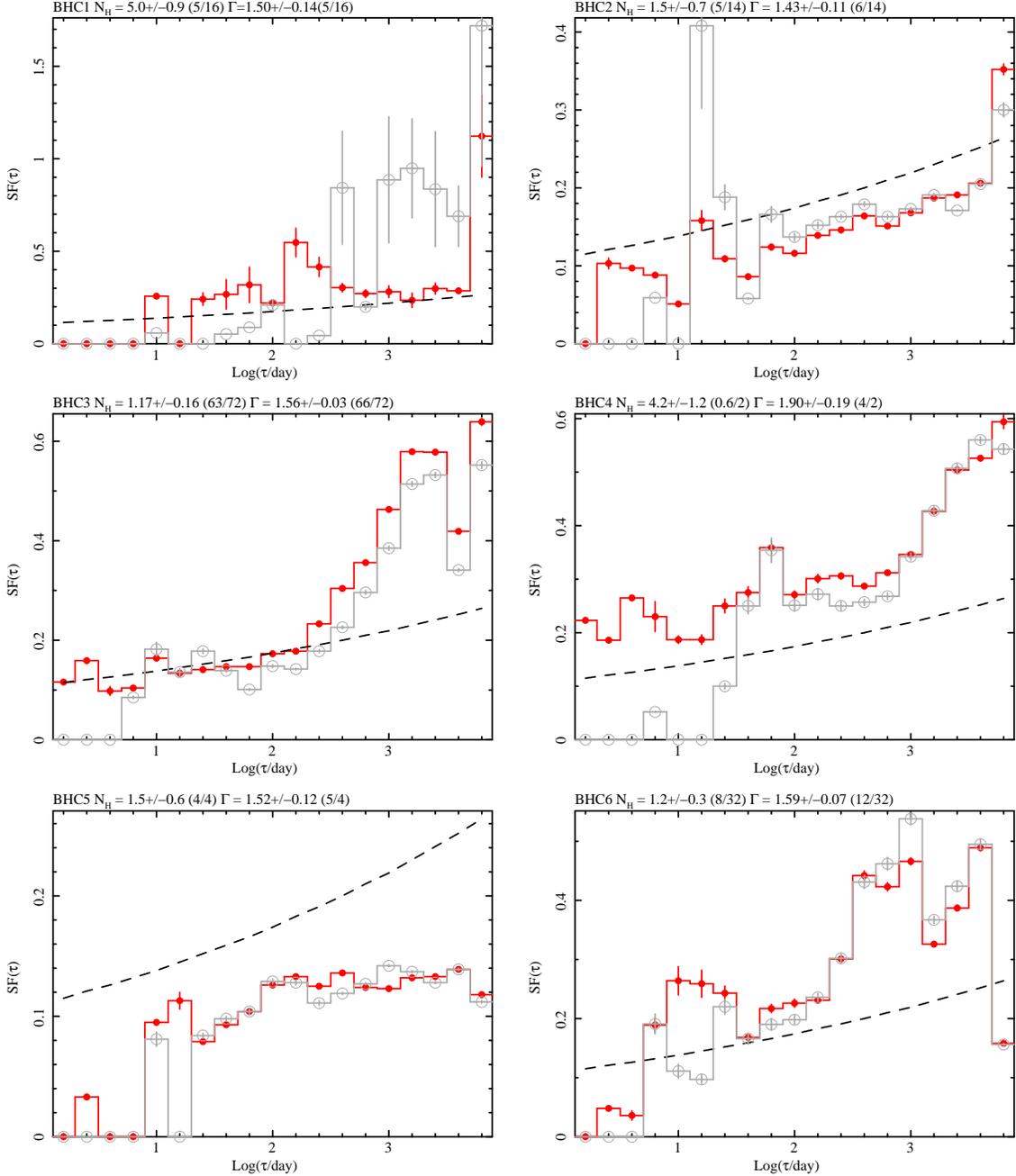}
\caption{Structure functions (SFs) for each BHC. In each case we provide the absorption (N$_{\rm H}$) and photon index ($\Gamma$) used to convert from 0.3--10 keV to 0.5--4.5 keV flux, along with the $\chi^2$/dof, obtained by fitting all ACIS observations with $>$200 counts; N$_{\rm H}$ is normalized to 10$^{21}$ H atom cm$^{-2}$. The SF with filled circles is drawn from ACIS and HRC observations, while the SF with open circles uses only ACIS data. The dashed line represents SF($\tau$) = 0.11$\tau^{0.10}$, our approximation of the Vagnetti et al. (2011) ensemble AGN SF. An extended, color version of this figure is available online.}\label{sfs}
\end{figure*}

\begin{table*}
\begin{center}
\caption{Details of the highest quality spectrum that indicates a  high luminosity hard state. We show the observation number, and net source counts. We then show the absorption ($N_{\rm H}$), spectral index ($\Gamma$), $\chi^2$/dof and 0.3--10 keV luminosity for the best fit power law model. Observations starting with zero are XMM-Newton observations; the others are from Chandra. Uncertainties are quoted at the 1$\sigma$ level.} \label{bestspec}
\renewcommand{\arraystretch}{.9}
\begin{tabular}{cccccccccccc}
\tableline\tableline
BHC & Obs & Cnts & $N_{\rm H}$/10$^{22}$ cm$^{-2}$ & $\Gamma$ & $\chi^2$/dof  & $L$ /10$^{37}$ \\
\tableline 
1 & 0109270101 & 3229 & 0.40$\pm$0.05 & 1.24$\pm$0.09 & 133/155 [0.89] & 27.1$\pm$1.1\\
2 & 303 & 704 & 0.15$\pm$0.10 & 1.5$\pm$0.2 & 33/29 [0.29] & 6.6$\pm$0.6\\
3 & 13827 & 5597 & 0.07 f & 1.50$\pm$0.04 & 161/182 [0.87] & 14.4$\pm$0.4\\
4 & 0690600401 & 2370 & 0.26$\pm$0.04 & 1.67$\pm$0.14 & 153/146 [0.32] & 6.6$\pm$0.3\\ 
5 & 13825 &1425 &0.21$\pm$0.09 & 1.68$\pm$0.18 & 8/11 [0.72] & 4.2$\pm$0.3\\
6 & 0112570101 & 10384 & 0.112$\pm$0.012 & 1.72$\pm$0.06 & 428/411 [0.26] & 7.30$\pm$0.17\\
7 & 0112570101 & 14523 & 0.110$\pm$0.010 & 1.76$\pm$0.04 & 519/490 [0.18] & 8.79$\pm$0.17\\
8 & 9524 & 352 & 0.07 f & 1.29$\pm$0.17 & 16/15 [0.37] & 17$\pm$3\\
9 & 1575 & 3097 & 0.10$\pm$0.03 & 1.73$\pm$0.09 & 106/109 [0.55] & 5.0$\pm$0.2\\
10 & 13827 & 1679 & 0.13$\pm$0.06 & 1.56$\pm$0.13 & 72/67 [0.32] & 3.6$\pm$0.2\\
11 & 303 & 704 & 0.10$\pm$0.09 & 1.43$\pm$0.18 & 58/45 [0.10] & 9.7$\pm$0.8\\
12 & 13825 & 2620 & 0.15$\pm$0.04 & 2.00$\pm$0.10 & 90/98 [0.71] & 5.2$\pm$0.2\\
13 & 14197 & 1593 & 0.12$\pm$0.05 & 1.63$\pm$0.14 & 72/64 [0.24] & 3.2$\pm$0.2\\
14 & 9521 & 183 & 0.07 f & 1.5$\pm$0.4 & 6/7 [0.56] & 10$\pm$3\\
15 & 14198 & 1860 & 0.13$\pm$0.05 & 1.56$\pm$0.12 & 75/74 [0.47] & 3.9$\pm$0.2\\
16 & 13825 & 4003 & 0.12$\pm$0.03 & 1.40$\pm$0.07 & 161/165 & 10.2$\pm$0.4\\
17 & 14197 & 1954 & 0.14$\pm$0.04 & 2.05$\pm$0.13 & 94/76 & 3.25$\pm$0.18\\
18 & 13825 & 4207 & 0.09$\pm$0.03 & 1.45$\pm$0.08 & 140/154 [0.78] & 9.0$\pm$0.4\\
19 & 0112570101 & 5032 & 0.41$\pm$0.03 & 1.89$\pm$0.08 & 210/226 [0.77] & 9.74$\pm$0.4\\
20 & 0402560901 & 20025 & 0.275$\pm$0.009 & 1.60$\pm$0.02 & 771/769 [0.44] & 45.0$\pm$0.4\\
21 & 13827 & 3847 & 0.22$\pm$0.05 & 1.73$\pm$0.09 & 127/142 [0.82] & 10.0$\pm$0.4\\
22 & 0112570101 & 10145 & 0.134$\pm$0.014 & 1.81$\pm$0.05 & 490/407 [3 E-3] & 8.2$\pm$0.2\\
23 & 14198 & 2508 & 0.13$\pm$0.04 & 1.64$\pm$0.10 & 111/95 [0.12] & 5.1$\pm$0.2\\
24 & 0112570101 & 6804 & 0.120$\pm$0.015 & 1.91$\pm$0.06 & 265/294 [0.89] & 5.13$\pm$0.15\\
25 & 0112570101 & 5116 & 0.108$\pm$0.019 & 1.50$\pm$0.06 & 254/236 [0.20] & 5.3$\pm$0.17\\
26 & 13825 & 1425 & 0.16$\pm$0.07 & 1.71$\pm$0.17 & 54/57 [0.60] & 3.3$\pm$0.2\\
27 & 14197 & 2099 & 0.14$\pm$0.04 & 1.83$\pm$0.12 & 69/75 [0.67] & 4.1$\pm$0.2\\
28 & 13826 & 635 & 0.07 f  & 1.73$\pm$0.13 & 23/27 [0.69] & 5.3$\pm$0.4\\
29 & 1577 & 1375 & 0.07 f & 1.33$\pm$0.08 & 79/58 [0.04] & 31$\pm$2\\
30 & 13299 & 274 & 0.07 f & 1.2$\pm$0.2 & 10/10 [0.42] & 9.9$\pm$1.6\\
31 & 0112570101 & 14338 & 0.081$\pm$0.009 & 1.64$\pm$0.04 & 450/488 [0.89] & 10.4$\pm$0.2\\
32 & 8184 & 521 & 0.07 f & 1.43$\pm$0.13 & 27/23 [0.25] & 13.4$\pm$1.5\\
33 & 13825 & 1411 & 0.14$\pm$0.08 & 1.81$\pm$0.17 & 57/60 [0.57] & 4.4$\pm$0.3\\
34 & 0112570101 & 5181 & 0.124$\pm$0.019 & 1.64$\pm$0.07 & 246/230 [0.22] & 7.0$\pm$0.2\\
35 & 14198 & 1646 & 0.18$\pm$0.08 & 1.75$\pm$0.16 & 85/69 [0.10] & 5.7$\pm$0.4\\
\tableline
\end{tabular}

\end{center}
\end{table*}

\subsection{Long term behavior}

We initially identified most of our BHCs from their long term ($\sim$13 year) light curves and spectral histories. They exhibited spectra consistent with the hard state ($\Gamma$ $<$2.1) at 0.3--10 keV luminosities $>$3$\times 10^{37}$ erg s$^{-1}$ in at least one observation.  For such spectra, we expect the 0.01--1000 keV luminosities to be $\gg$3$\times 10^{37}$ erg s$^{-1}$, our limit for neutron star hard states following \citet{gladstone07}. 

We present the $\sim$13 year, corrected,  0.3--10 keV luminosity lightcurve for each BHC in Fig.~\ref{lcs}, along with its spectral history. The top panel in each case shows the lightcurve, with ACIS and HRC observations represented by circles and crosses respectively. The bottom panel shows $\Gamma$ for each ACIS observation; sufficiently bright observations have freely fitted $\Gamma$, while faint observations have $\Gamma$ fixed to the mean value; this approach should be valid because we expect the faint observations to be in the hard state also, with similar a emission spectrum.

Figure~\ref{hist} shows histograms of the number of observation pairs with separation $\tau$ for the most frequently observed BHC, BHC13. The histogram with filled circles includes ACIS and HRC observations, while the open circle histogram contains only ACIS observations. The observation pairs were logarithmically binned by $\tau$, with a bin width of 0.2 dex. 

We present the 0.5--4.5 keV structure function for each of our BHCs in Fig.~\ref{sfs}, with same scheme as for Fig.~\ref{hist}. The ensemble SF for typical AGN obtained by Vagnetti et al. (2011) is represented by a dashed line in each panel: SF($\tau$) = 0.11$\tau^{0.10\pm0.01}$.  We included an ACIS only SF for each source in addition to an ACIS + HRC SF because faint HRC observations add extremely large noise components for some BHCs, suppressing the variation; BHCs 8 and 35 only show significant excess variability over typical AGN in their  ACIS only SFs.

We find that 28  out of 35 BHCs exhibit SFs with significantly more variation than the Vagnetti et al. (2011) ensemble AGN SF ($>$4$\sigma$ excess in at least one $\tau$ channel, $>$3$\sigma$ excess in at least 2 channels, or $>$2$\sigma$ excess in at least 3 channels  over the 3$\sigma$ limit for the AGN SF, assumed to be 0.11$\tau^{0.13}$). This distinction is important, because AGN and XBs often have very similar spectra. The   X-ray sources with SFs exhibiting similar or less variation than the ensemble AGN SF (BHCs 2, 5, 7, 21, 23, 24, 25,  and 34) had mean 0.3--10 keV fluxes $\ga$5.5$\times 10^{-13}$ erg cm $^{-2}$ s$^{-1}$.

We have found 24 X-ray sources  in our  observations  that exhibited mean 0.3--10 keV fluxes $\ga$5.5$\times$10$^{-13}$ erg cm$^{-2}$ s$^{-1}$ over the $\sim$13 years of Chandra monitoring; this includes 8 GC XBs and 1 possible supernova remnant. The 0.5--10 keV flux distribution of AGN found by \citep{georgakakis08} predicts 0.6 AGN  at this flux level within our observed region (approximated by a circle with radius $\sim$20$'$ or $\sim$4 kpc). Since the observed number of unidentified   X-ray sources at these flux levels is larger than the predicted AGN number by a factor $\sim$25, we conclude that they are likely XBs. 

\subsection{Spectral analysis}

After the initial identification of candidates, we sought the observations that best represented the BHC case for each object; these were the highest quality spectra that appeared to exhibit high luminosity hard states. Table~\ref{bestspec} provides the details of these observations.  For each BHC we give the observation number, and net sources counts. We then give the absorption ($N_{\rm H}$), photon index ($\Gamma$), $\chi^2$/dof and 0.3--10 keV luminosity for the best fit absorbed power law model. Observations 303--13827 are Chandra ACIS observations, while 0109270101, 0112570101, and 0402560901 are XMM-Newton observations.

We note that the spectrum of BHC 22 rejected simple absorbed power law models. It is well described by an absorbed  disk blackbody + power law model with $T_{\rm in}$ = 1.97$_{-0.3}^{+0.19}$ keV, and $\Gamma$ = 2.16$_{-0.19}^{+0.4}$, with the power law contributing 61$\pm$12\% of the 8.0$\pm$1.1 $\times 10^{37}$ erg s$^{-1}$ 0.3--10 keV luminosity; it is consistent with either of two canonical black hole states:  the hard state, or a steep power law state similar to GRS 1915+105 (McClintock \& Remillard, 2006).

We applied a disk blackbody + blackbody emission model to our BHC spectra. The disk blackbody components were generally better constrained than the blackbody components, particularly for spectra with relatively few net source counts; this is simply because there are fewer high energy photons than low energy photons and the disk blackbody is cooler than the blackbody. Our best fits yielded blackbody contributions of 65--99\%  to the 2--10 keV flux, in stark contrast to the neutron star systems where the disk blackbody dominates. The blackbody component dominates the 2--10 keV flux most when the disk blackbody temperature is lowest.
Four of our BHCs only exhibited apparent high luminosity hard states during 5 ks ACIS observations, and we were unable to constrain double thermal models for the resulting spectra. However, future deep observations may strengthen their BHC cases. 

The 31 BHCs that we did fit with the double thermal model yielded k$T_{\rm in}$ 0.6--28$\sigma$ outside the NS parameter space; the combined $\chi^2$ was 212, for 30 dof (probability $\sim$3$\times 10^{-29}$). Clearly our BHCs are inconsistent with being soft NS XBs.

\begin{table*}
\begin{center}
\caption{ Best fit parameters for the blackbody + disk blackbody model. We provide the absorption , blackbody temperature, 2--10 keV blackbody luminosity, disk blackbody temperature, and 2--10 keV disk blackbody luminosity. Finally we give the blackbody contribution to the total 2--10 keV luminosity. Uncertainties are quoted at the 1$\sigma$ level. The last column states whether the BHC identification is strong (S) or plausible (P)} \label{twotempfits}
\renewcommand{\arraystretch}{.9}
\begin{tabular}{ccccccccccccc}
\tableline\tableline
BHC & $N_{\rm H}$ / 10$^{22}$ cm$^{-2}$ &  k$T_{\rm BB}$ / keV &  $L_{\rm BB}$/10$^{37}$ & k$T_{\rm DBB}$ / keV & $L_{\rm DBB}$/10$^{37}$ & $L_{\rm BB}/L_{\rm TOT}$ &BH  \\
\tableline 
1 & 0.34$\pm$0.07 & 2.00$_{-0.19}^{+0.6}$  & 21$\pm$2 & 0.88$_{-0.16}^{+0.4}$  &  7.6$\pm$0.5 & 0.73$\pm$0.09 & P\\
2 & 0.07f & 1.12$_{-0.11}^{+0.4}$  & 3.0$\pm$0.5 & 0.50$_{-0.10}^{+0.3}$  &  0.8$\pm$0.4 & 0.8$\pm$0.2 & P\\
3 & 0.07f & 1.18$_{-0.08}^{+0.13}$  & 7.3$\pm$0.3 & 0.49$_{-0.06}^{+0.08}$  &  0.8$\pm$0.3 & 0.90$\pm$0.06 & S\\
4 & 0.19$\pm$0.06 & 1.50$_{-0.19}^{+1.0}$  & 3.9$\pm$0.8 & 0.62$_{-0.11}^{+0.2}$  &  0.8$\pm$0.4 & 0.8$\pm$0.2 & P\\
5 & 0.07f & 1.3$_{-0.3}^{+25.0}$  & 2.0$\pm$0.7 & 0.7$_{-0.2}^{+0.5}$  &  0.6$\pm$0.3 & 0.8$\pm$0.4 & P\\
6 & 0.07f & 1.34$_{-0.05}^{+0.06}$  & 3.67$\pm$0.15 & 0.510$_{-0.013}^{+0.02}$  &  0.42$\pm$0.05 & 0.90$\pm$0.05 & S\\
7 & 0.07f & 1.47$_{-0.06}^{+0.06}$  & 4.59$\pm$0.13 & 0.51$_{-0.02}^{+0.03}$  &  0.54$\pm$0.06 & 0.89$\pm$0.04 &S \\
8 & --- & --- & --- & --- & --- & ---& P  \\
9 & 0.07f &  0.93$_{-0.06}^{+0.08}$  & 2.14$\pm$0.12 & 0.38$_{0.04}^{+0.04}$  &  0.14$\pm$0.05 & 0.94$\pm$0.08 & S\\
10 & 0.07f & 2.4$_{-0.9}^{+12.5}$  & 2.3$\pm$0.6 & 0.86$_{-0.19}^{+0.2}$  &  0.7$\pm$0.3 & 0.8$\pm$0.3 & P\\
11 & 0.07f & 1.11$_{-0.11}^{+0.4}$  & 3.0$\pm$0.5 & 0.50$_{-0.10}^{+0.3}$  &  0.8$\pm$0.2 & 0.80$\pm$0.18 & P\\
12 & 0.07f & 1.24$_{-0.19}^{+0.7}$  & 1.74$\pm$0.17 & 0.59$_{-0.08}^{+0.10}$  &  0.7$\pm$0.3 & 0.72$\pm$0.12 & S\\
13 & 0.07f & 1.5$_{-0.3}^{+1.4}$  & 1.8$\pm$0.4 & 0.67$_{-0.12}^{+0.2}$  &  0.5$\pm$0.3 & 0.8$\pm$0.2 & P\\
14 & --- & --- & --- & --- & --- & ---& P \\
15 & 0.07f & 1.12$_{-0.11}^{+0.4}$  & 1.6$\pm$0.2 & 0.56$_{-0.11}^{+0.3}$  &  0.6$\pm$0.3 & 0.74$\pm$0.17 & P\\
16 & 0.07f & 2.2$_{-0.5}^{+2.4}$  & 4.5$\pm$0.5 & 1.03$_{-0.19}^{+0.2}$  &  2.4$\pm$0.8 & 0.65$\pm$0.11 & P\\
17 & 0.07f & 0.96$_{-0.09}^{+0.13}$  & 1.17$\pm$0.09 & 0.43$_{-0.05}^{+0.06}$  &  0.17$\pm$0.07 & 0.87$\pm$0.10 & S\\
18 & 0.07f & 1.05$_{-0.05}^{+0.08}$  & 4.44$\pm$0.19 & 0.41$_{-0.04}^{+0.06}$  &  0.22$\pm$0.08 & 0.95$\pm$0.06 & S\\
19 & 0.26$\pm$0.04 & 1.7$_{-0.3}^{+1.1}$  & 3.6$\pm$0.5 & 0.83$_{-0.19}^{+0.2}$  &  1.7$\pm$0.5 & 0.68$\pm$0.13 & P\\
20 & 0.173$\pm$0.010 & 1.61$_{-0.17}^{+0.19}$  & 18.1$\pm$1.7 & 0.83$_{-0.10}^{+0.12}$  &  6.9$\pm$1.9 & 0.7$\pm$0.2& S\\
21 & 0.1$\pm$0.05 & 1.7$_{-0.3}^{+0.8}$  & 4.1$\pm$0.5 & 0.77$_{-0.13}^{+0.2}$  &  1.9$\pm$0.7 & 0.68$\pm$0.12 & P\\
22 & 0.07f & 1.31$_{-0.05}^{+0.06}$  & 3.71$\pm$0.13 & 0.55$_{-0.03}^{+0.03}$  &  0.57$\pm$0.08 & 0.87$\pm$0.04 & S\\
23 & 0.07f & 0.99$_{-0.07}^{+0.13}$  & 2.6$\pm$0.5 & 0.45$_{-0.06}^{+0.13}$  &  0.25$\pm$0.13 & 0.9$\pm$0.2 & S\\
24 & 0.07f & 1.12$_{-0.05}^{+0.06}$  & 2.12$\pm$0.09 & 0.43$_{-0.02}^{+0.03}$  &  0.19$\pm$0.04 & 0.92$\pm$0.05 & S\\
25 & 0.07f & 1.56$_{-0.08}^{+0.10}$  & 3.22$\pm$0.13 & 0.61$_{-0.04}^{+0.04}$  &  0.39$\pm$0.07 & 0.89$\pm$0.05 & S\\
26 & 0.07f & 2.3$_{-1.6}^{+20.0}$  & 2.1$\pm$0.6 & 0.78$_{-0.16}^{+0.2}$  &  0.5$\pm$0.2 & 0.8$\pm$0.3 & P\\
27 & 0.07f & 0.94$_{-0.08}^{+0.3}$  & 1.56$\pm$0.14 & 0.46$_{-0.08}^{+0.13}$  &  0.28$\pm$0.15 & 0.85$\pm$0.12 & S \\
28 & 0.07f & 1.4$_{-0.3}^{+1.0}$  & 2.3$\pm$0.5 & 0.63$_{-0.12}^{+0.2}$  &  0.9$\pm$0.4 & 0.71$\pm$0.2 & P \\
29 & 0.07f & 1.00$_{-0.05}^{+0.07}$  & 16.4$\pm$0.9 & 0.29$_{-0.04}^{+0.06}$  &  0.22$\pm$0.11 & 0.99$\pm$0.08 & S \\
30 & --- & --- & --- & --- & --- & ---& P \\
31 & 0.07f & 1.36$_{-0.04}^{+0.05}$  & 5.72$\pm$0.15 & 0.45$_{-0.02}^{+0.02}$  &  0.4$\pm$0.05 & 0.93$\pm$0.03 & S \\
32 & --- & --- & --- & --- & --- & ---& P   \\
33 & 0.07f & 1.15$_{-0.12}^{+0.3}$  & 1.87$\pm$0.17 & 0.52$_{-0.08}^{+0.12}$  &  0.38$\pm$0.17 & 0.83$\pm$0.12& S\\
34 & 0.07f & 1.56$_{-0.11}^{+0.13}$  & 3.7$\pm$0.3 & 0.63$_{-0.04}^{+0.04}$  &  0.67$\pm$0.12 & 0.85$\pm$0.08 & S\\
35 & 0.07f & 1.44$_{-0.19}^{+1.2}$  & 2.4$\pm$0.4 & 0.68$_{-0.11}^{+0.18}$  &  0.8$\pm$0.3 & 0.75$\pm$0.17 & P\\

\tableline
\end{tabular}

\end{center}
\end{table*}


\begin{figure}
\epsscale{1.1}
\plotone{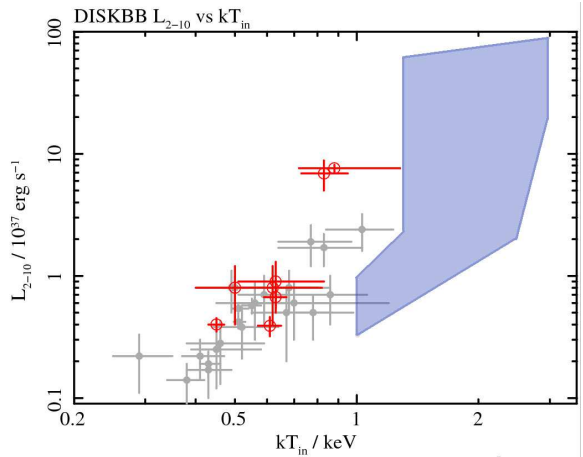}
\caption{ Comparison of disk blackbody 2--10 keV luminosity  vs k$T_{\rm in}$ for the best fit double thermal models to our BHC spectra. The  shaded region encompasses the parameter space inhabited by  hundreds of spectral fits for Aql X-1, 4U 1608$-$52, XTE J1701$-$462 and GX 17+2; together, they represent all types of NS LMXB behavior (Lin et al. 2007, 2009, 2012). Filled circles represent field BHCs, and open circles represent GC BHCs. We note that the disk blackbody component contributes only $\sim$1--35\% of the total 2--10 keV flux for our BHCs, but dominates the spectra studied by Lin et al. (2007, 2009, 2012). The probability that the BHCs all lie within the shaded region is $\sim$3$\times 10^{-29}$.}\label{dbbvl}
\end{figure}

In Figs. \ref{dbbvl} and \ref{bbvl} we plot the 2.0--10 keV luminosity vs. temperature for the disk blackbody and blackbody components of the best fit double thermal models for each of our BHCs. We overlay the parameter space inhabited by the neutron star systems  observed by Lin et al. (2007, 2009, 2012) for spectral states where Comptonization is not required in their models, i.e. the atoll soft states and Z-source normal and flaring branches. We note that Lin et al. quote luminosities from broader energy ranges; however, our disk blackbody components make a negligible contribution above 10 keV, while the luminosities of our blackbody components are typically $\sim$10\% higher in the 2.0--20 keV band than the 2.0--10 keV band; hence our observed 2.0--10 keV luminosity for our BHCs provide a fair comparison with the neutron star systems observed by Lin et al. (2007, 2009, 2012).  

As predicted, our BHCs inhabit a separate region of the disk blackbody  parameter space to the NS systems studied by Lin et al. (2007, 2009, 2012), although some sources are consistent with  NS values for $T_{\rm in}$ within 3$\sigma$. We also see a correlation between the 2--10 keV luminosity and temperature of the disk blackbody component; this is due to  lower disk temperatures resulting in lesser contributions to the 2--10 keV flux.   The BHCs are scattered more widely in the blackbody space, but are also systematically offset from the NS systems.



\begin{figure}
\epsscale{1.1}
\plotone{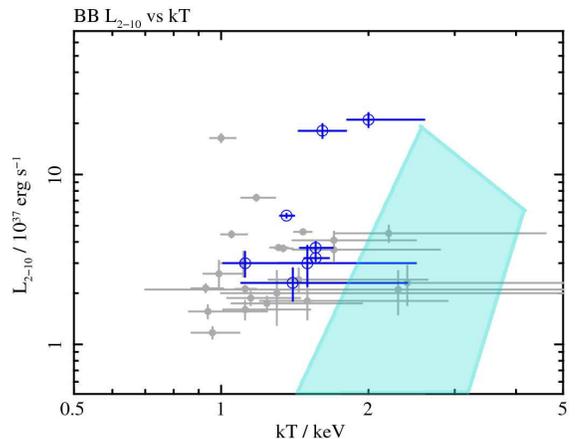}
\caption{ Comparison of blackbody 2--10 keV luminosity  vs k$T$ for the best fit double thermal models to our BHC spectra.  Filled circles represent field BHCs, and open circles represent GC BHCs. The  shaded region encompasses the parameter space inhabited by hundreds of spectral fits for Aql X-1, 4U 1608$-$52, XTE J1701$-$462 and GX 17+2, which together exhibit every type of NS LMXB behavior (Lin et al. 2007, 2009, 2012). We note that the blackbody component contributes $\sim$65--99\% of the total 2--10 keV flux for our BHCS, but less than 50\% for the NS sources studied by Lin et al. (2007, 2009, 2012).  }\label{bbvl}
\end{figure}


Remembering that  Lin et al. (2009, 2012) failed to fit the horizontal branches of  XTE J1701$-$462 and GX 17+2 with their double thermal model, we also modeled our brightest BHCs ($L_{0.3-10}$ $>$ 10$^{38}$ erg s$^{-1}$) with the disk blacbkody + blackbody + Comptonization  emission model that they employed for the horizontal branch spectra. \citet{lin12} used a cut-off power law to model the Comptonization in GX 17+2  over the 2.9--60 keV band; they required simultaneous fitting of several spectra to constrain the parameters of the cut-off power law, which yielded $\Gamma$ = 1.40$\pm$0.14, and a cut-off energy of 9.9$\pm$1.0 keV. We therefore modeled the 0.3--10 keV XMM-Netwon pn spectra, or 0.3--7.0 keV Chandra ACIS  spectra of our brightest BHCs with a blackbody, disk blackbody and a power law with $\Gamma$ fixed to 1.4.  \citet{lin12} found that the Compotonized component contributed $\sim$10--50\% of the total flux on the horizontal branch of GX 17+2; we fixed the flux of the power law component in our fits to $\sim$30\% of the total 2--10 keV flux.  

We fitted the spectra of BHCs 1, 3, 16, 20, 21, and 31 in this way; the results are summarized in Table~\ref{bdp}. We find that this three component model results in lower temperatures for both thermal components than the  double thermal model fits to the same spectra. I.e., the addition of the power law component drives our BHCs away from the NS parameter space. Increasing  the power law contribution to 50\% of the flux decreased the temperatures further.  Since none of the spectra from our brightest BHCs have three component fits that are consistent with the fits that successfully described the horizontal branch (Lin et al., 2009, 2012), we conclude that none of our BHCs are Z-sources.

\section{Summary and conclusions}

We have identified 26 new black hole candidates in the central region of M31, using their structure functions or luminosities to identify them as X-ray binaries, and their high luminosity hard state spectra to classify them as BHCs. Of these, 12 are strong candidates, and 14 are plausible candidates that may benefit from further observations.  We were previously limited to identifying BHCs in globular clusters, due to the similarities between XB and AGN spectra. This brings the total number of BHCs within 20$'$ of M31$^*$ identified by their high luminosity hard states to 35.
\begin{table}[b]
\begin{center}
\caption{ Best fit blackbody and disk blackbody temperatures when our brightest BHCs are modeled with a blackbody, a disk blackbody, and a Comptonized component represented by a power law with $\Gamma$ = 1.4 that contributes $\sim$30\% of the 2--10 keV flux. This is for comparison with the spectral modeling conducted by Lin et al. (2012) for the horizontal branch of GX 17+2} \label{bdp}
\renewcommand{\arraystretch}{.9}
\begin{tabular}{ccccc}
\tableline\tableline
BHC & $N_{\rm H}$ / 10$^{22}$ cm$^{-2}$ &  k$T_{\rm BB}$ / keV &   k$T_{\rm DBB}$ / keV   \\
\tableline 
1 & 0.35 f & 1.99$_{-0.4}^{+1.95}$  & 1$_{-0.3}^{+1.1}$ \\
3 & 0.07 f & 1.07$_{-0.11}^{+0.19}$  & 0.42$_{-0.1}^{+0.15}$ \\
16 & 0.07 f & 1.44$_{-0.3}^{+...}$  & 0.8$_{-0.3}^{+0.8}$ \\
20 & 0.197$\pm$0.011 & 1.29$_{-0.13}^{+0.2}$  & 0.73$_{-0.13}^{+0.19}$ \\
21 & 0.11$\pm$0.07 & 1.6$_{-0.4}^{+...}$  & 0.7$_{-0.2}^{+0.4}$ \\
31 & 0.07 f & 1.27$_{-0.08}^{+0.09}$  & 0.43$_{-0.03}^{+0.03}$ \\
\tableline
\end{tabular}

\end{center}
\end{table}

The structure functions of most of our BHCs reveal them to be substantially more variable than typical AGN (as measured by Vagnetti et al., 2011) over a wide range of time scales. Those BHCs with comparable or less variability than the AGN have 0.5--10 keV luminosities matched by $\sim$0.6 AGN within the observed region, according to \citet{georgakakis08}. It is therefore unlikely that our new  BHCs are AGN.

We have found that our BHC spectra exist in a separate parameter space to Galactic neutron star systems when we compare our best fits for absorbed blackbody + disk blackbody emission models with the systems studied by Lin et al. (2007, 2009, 2012). This is expected because   this double thermal model fails to fit hard state spectra;   the disk blackbody component is forced  low temperatures in order to provide the low energy flux. Indeed, the probability that our BHCs all lie in the NS parameter space when fitted with the double-thermal model is just $\sim$3$\times 10^{-29}$.
Seven of our BHCs (20\% of our total sample) were found within 100$''$ of the M31 nucleus, lending support to the hypothesis that the M31 bulge is sufficiently dense to form a significant number of XBs dynamically, as seen in globular clusters; since the stellar velocities in the M31 bulge are considerably higher than in GCs, surviving XBs are expected to have short periods and are likely to contain black hole accretors  \citep{voss07}.



\section*{Acknowledgments}
We thank the referee for their thoughtful comments. We thank Z. Li for merging the Chandra data.  This research has made use of data obtained from the Chandra data archive, and software provided by the Chandra X-ray Center (CXC). We also include analysis of data from XMM-Newton, an ESA science mission with instruments and and contributions directly funded by ESA member states and the US (NASA).
R.B. is funded by Chandra grants GO0-11106X and GO1-12109X, along with  HST grants GO-11833 and GO-12014.  M.R.G. and S.S.M are  partially supported by NASA grant NAS-03060.




{\it Facilities:} \facility{CXO (ASIS)} \facility{CXO (HRC)}.

\clearpage



\addtocounter{figure}{-5}
\begin{figure*}
\epsscale{1}
\plotone{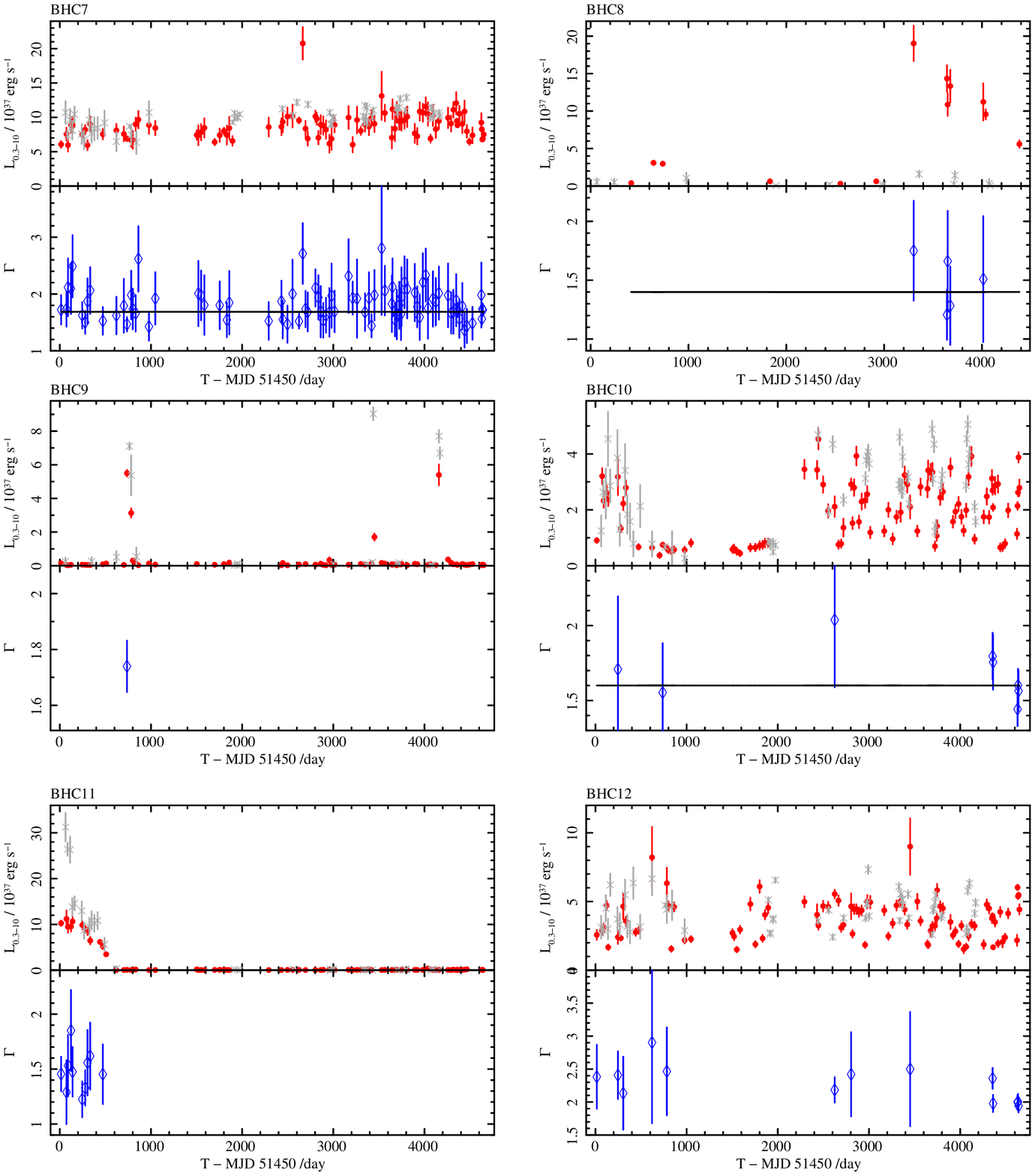}
\caption{continued}
\end{figure*}

\addtocounter{figure}{-1}
\begin{figure*}
\epsscale{1}
\plotone{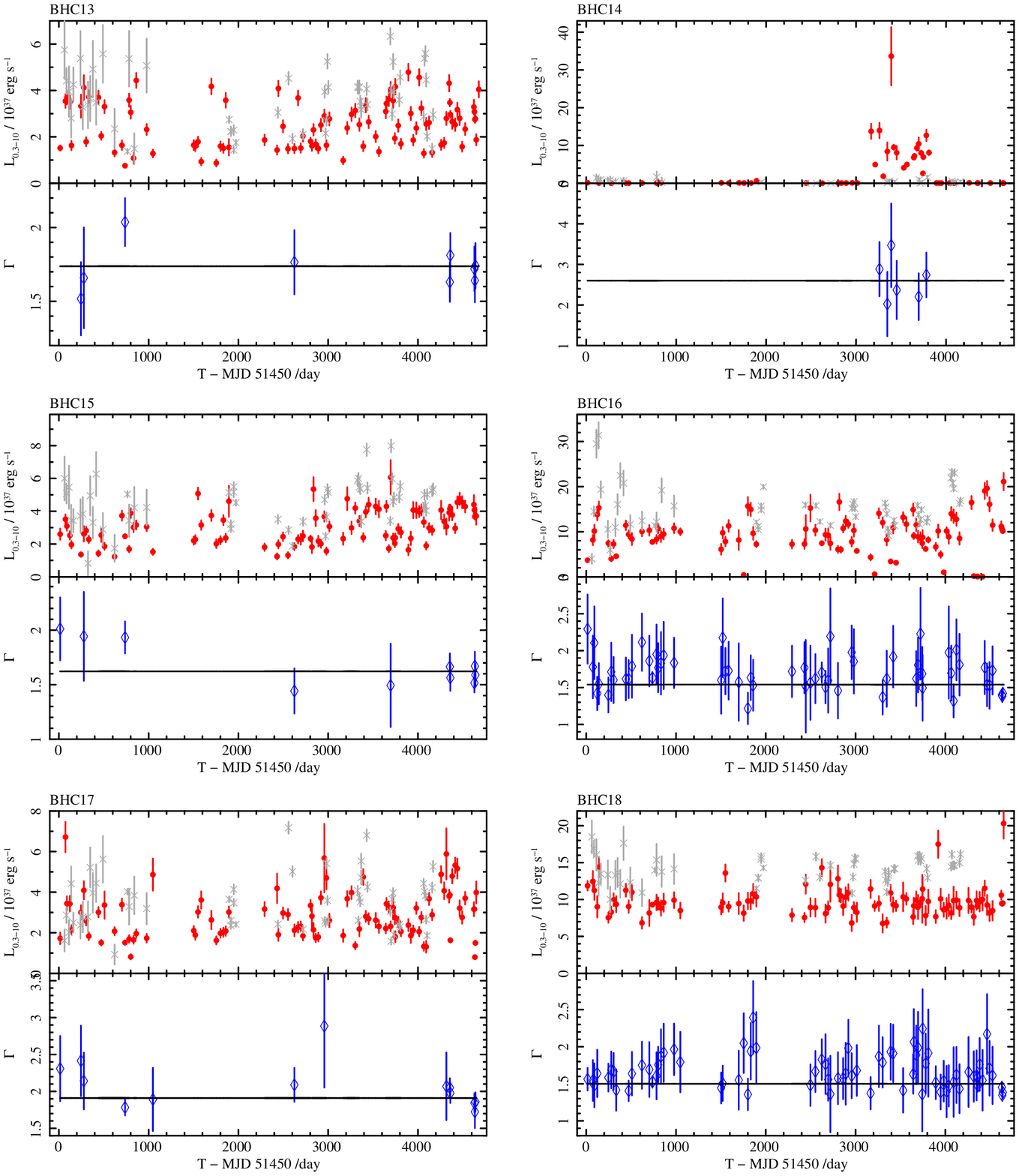}
\caption{continued}
\end{figure*}

\addtocounter{figure}{-1}
\begin{figure*}
\epsscale{1}
\plotone{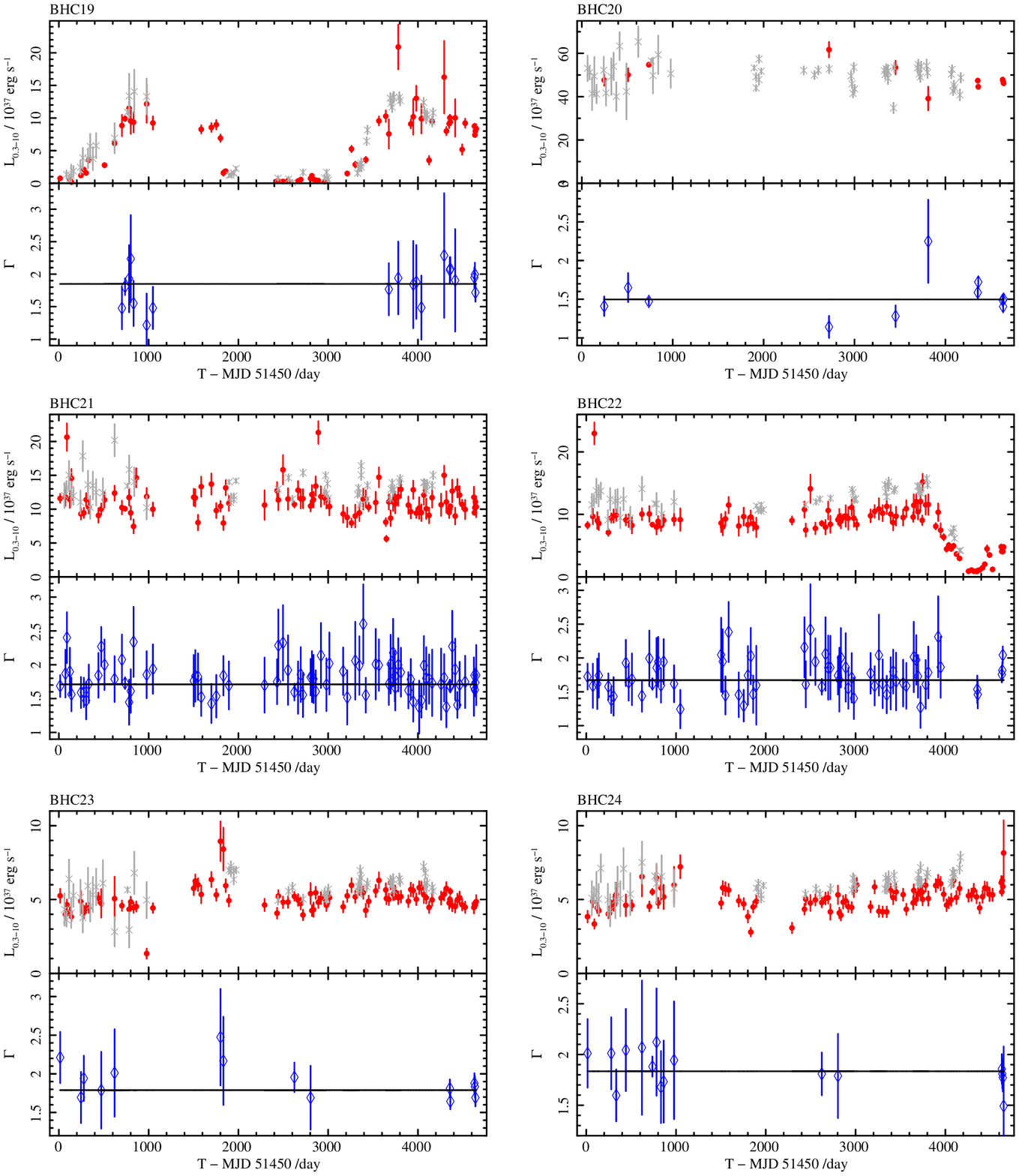}
\caption{continued}
\end{figure*}

\addtocounter{figure}{-1}
\begin{figure*}
\epsscale{1}
\plotone{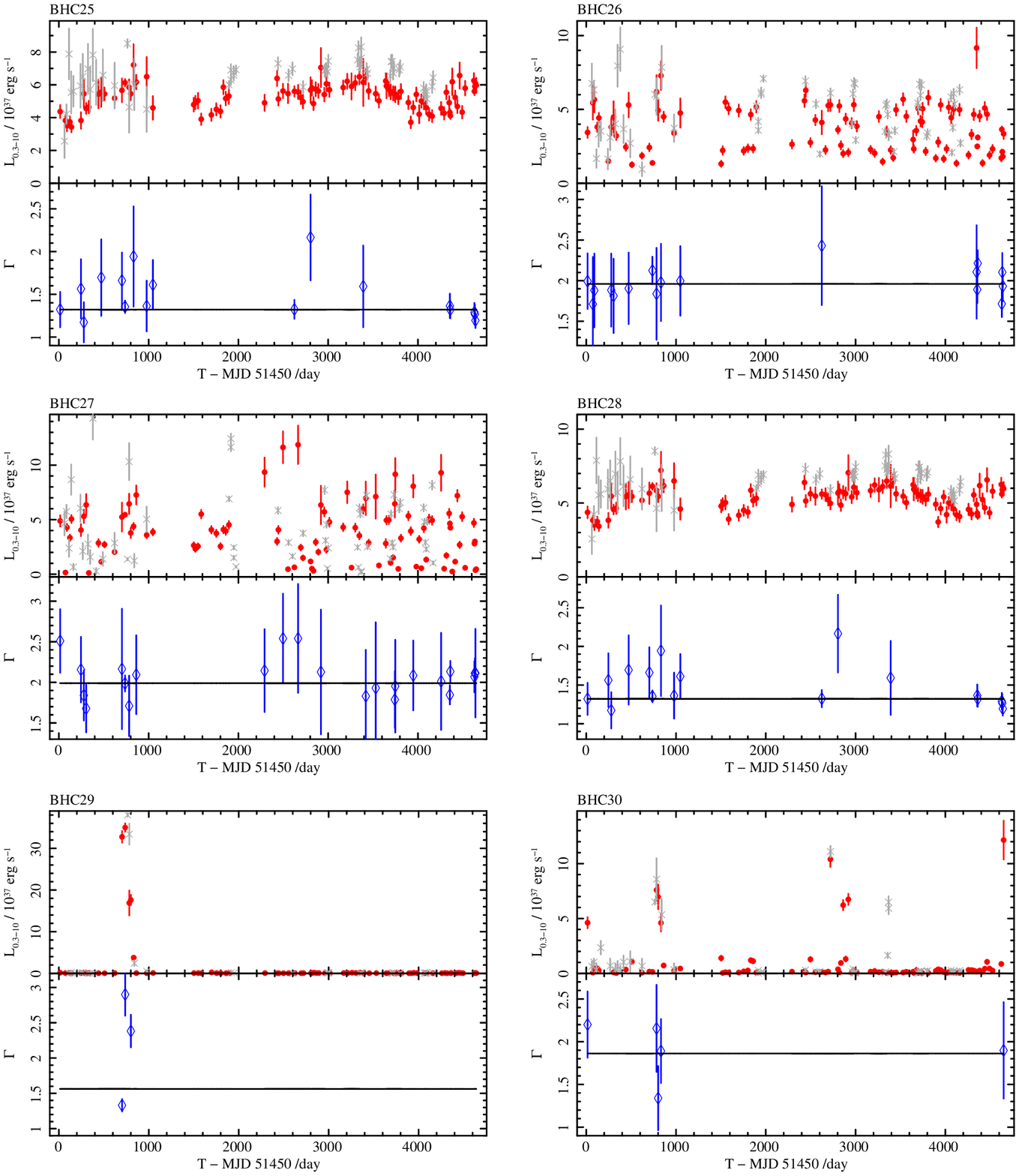}
\caption{continued}
\end{figure*}

\addtocounter{figure}{-1}
\begin{figure*}
\epsscale{1}
\plotone{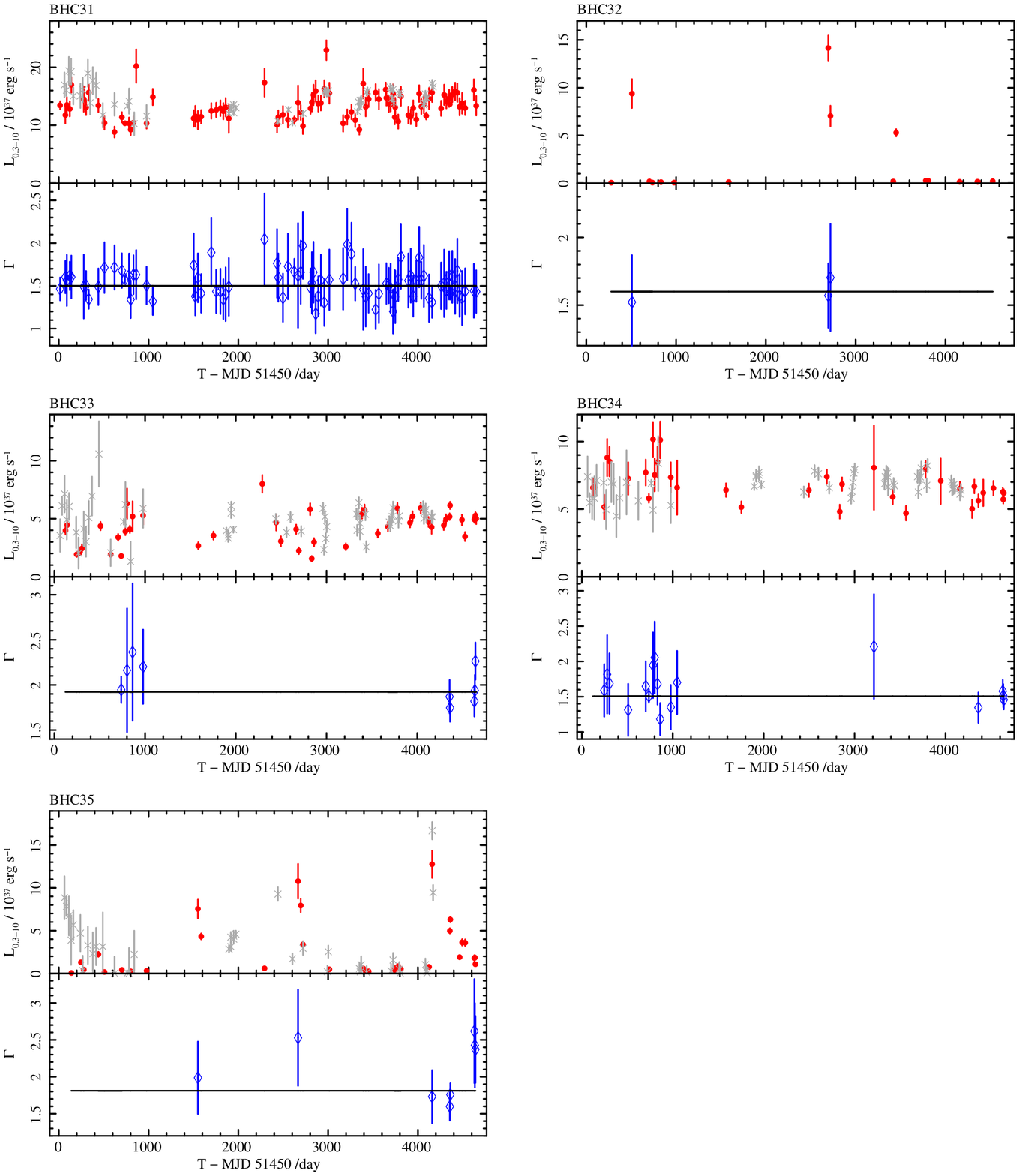}
\caption{continued}
\end{figure*}


\addtocounter{figure}{+1}
\begin{figure*}
\epsscale{1}
\plotone{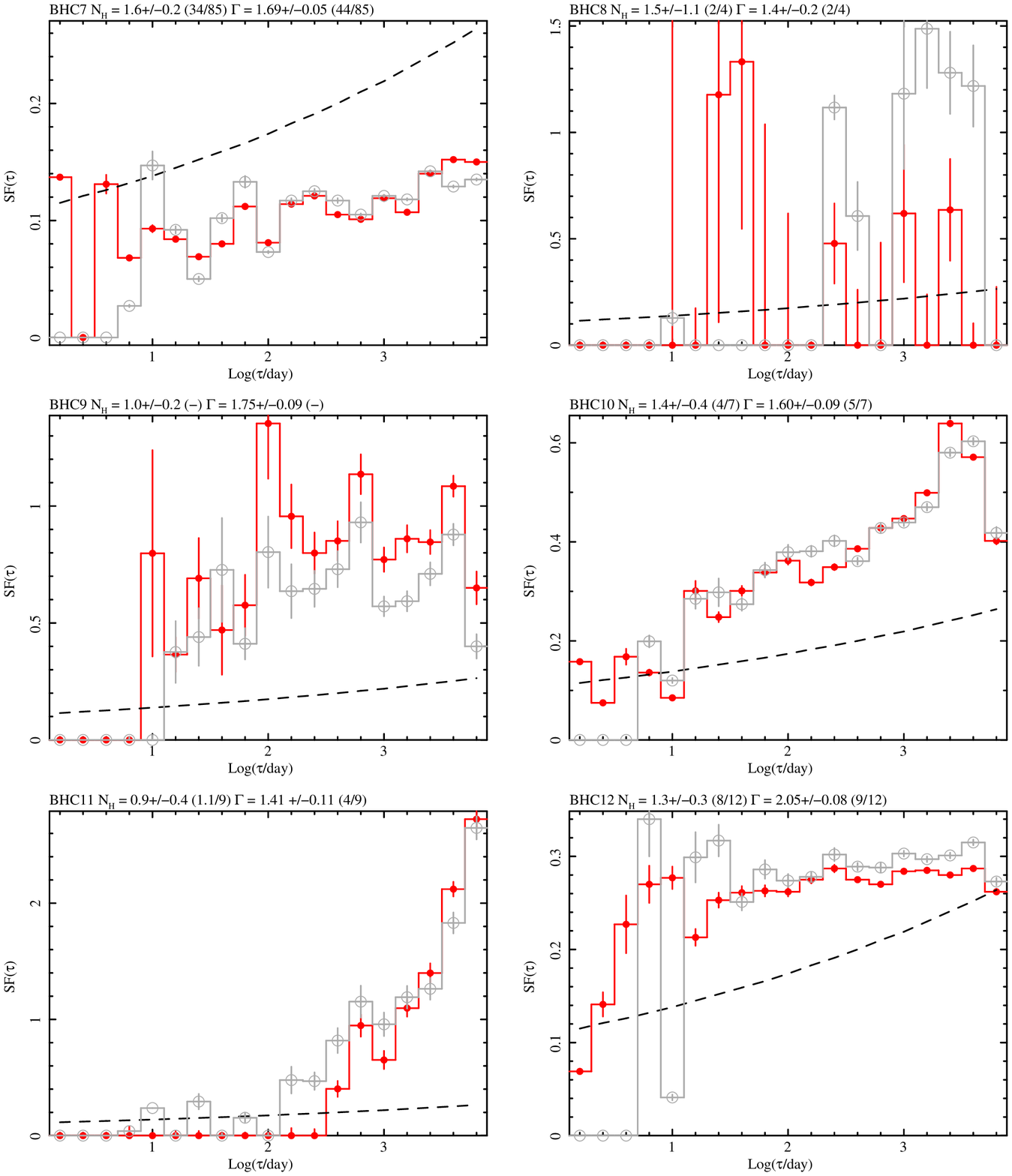}
\caption{continued}
\end{figure*}

\addtocounter{figure}{-1}
\begin{figure*}
\epsscale{1}
\plotone{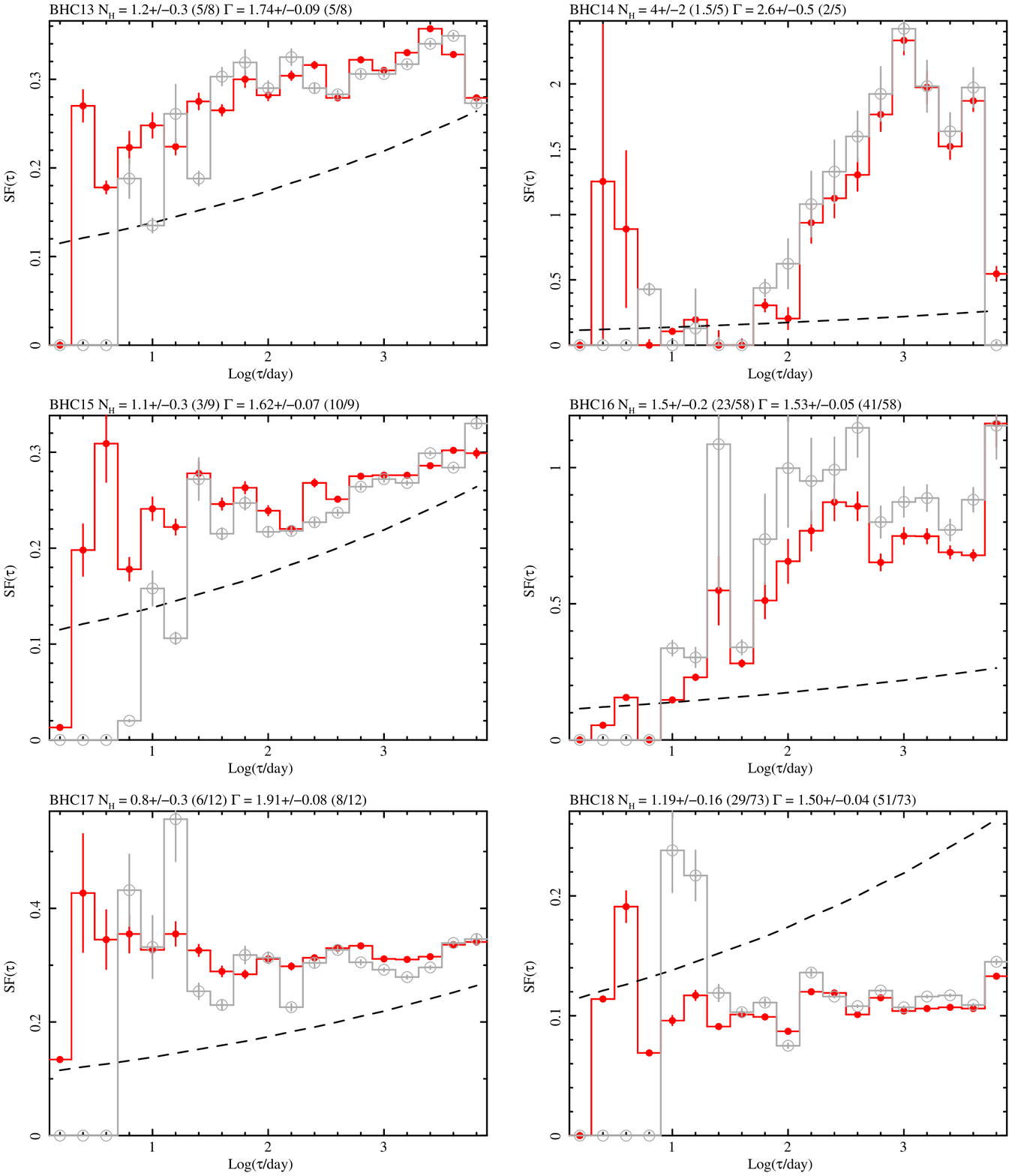}
\caption{continued}
\end{figure*}

\addtocounter{figure}{-1}
\begin{figure*}
\epsscale{1}
\plotone{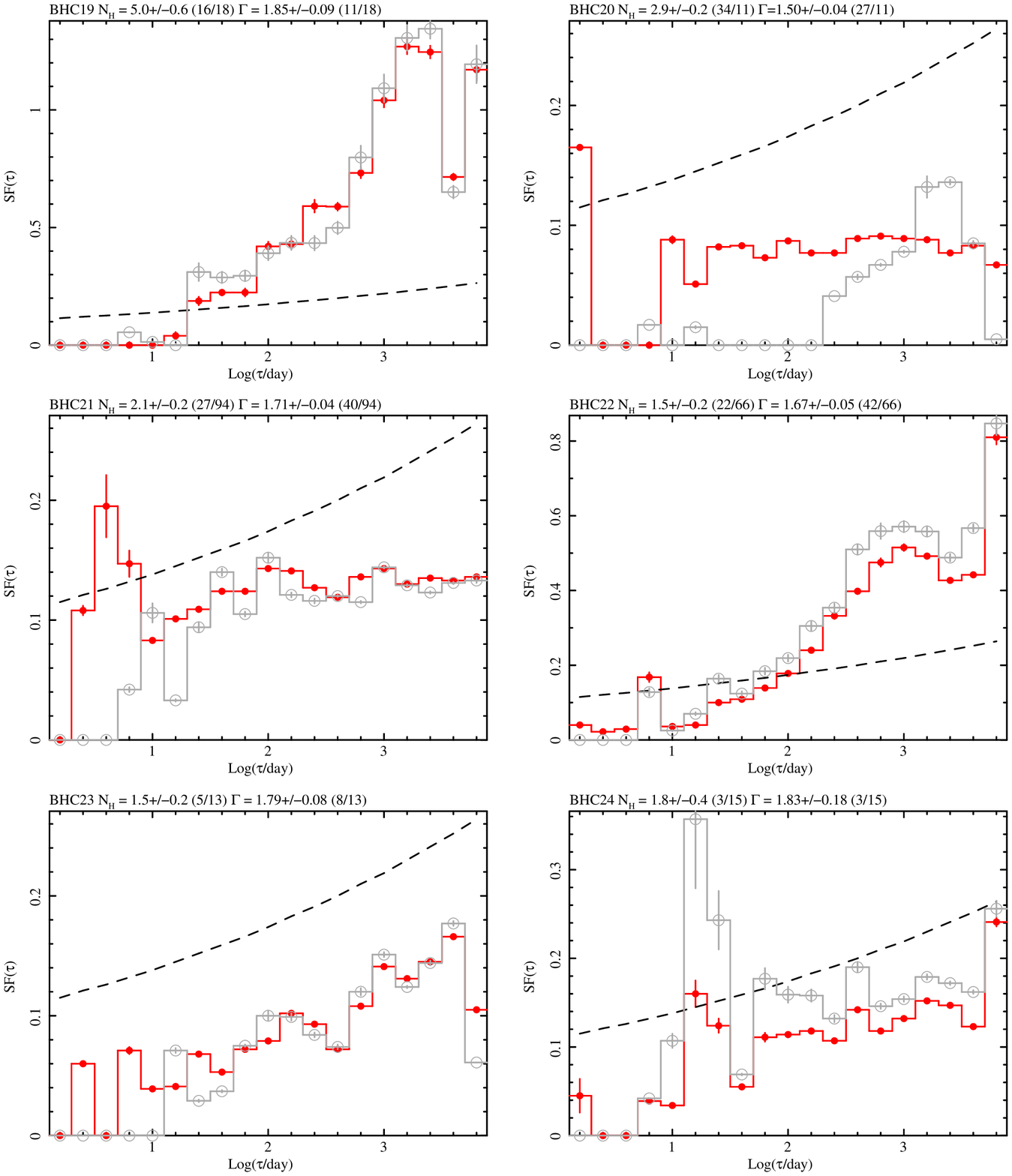}
\caption{continued}
\end{figure*}

\addtocounter{figure}{-1}
\begin{figure*}
\epsscale{1}
\plotone{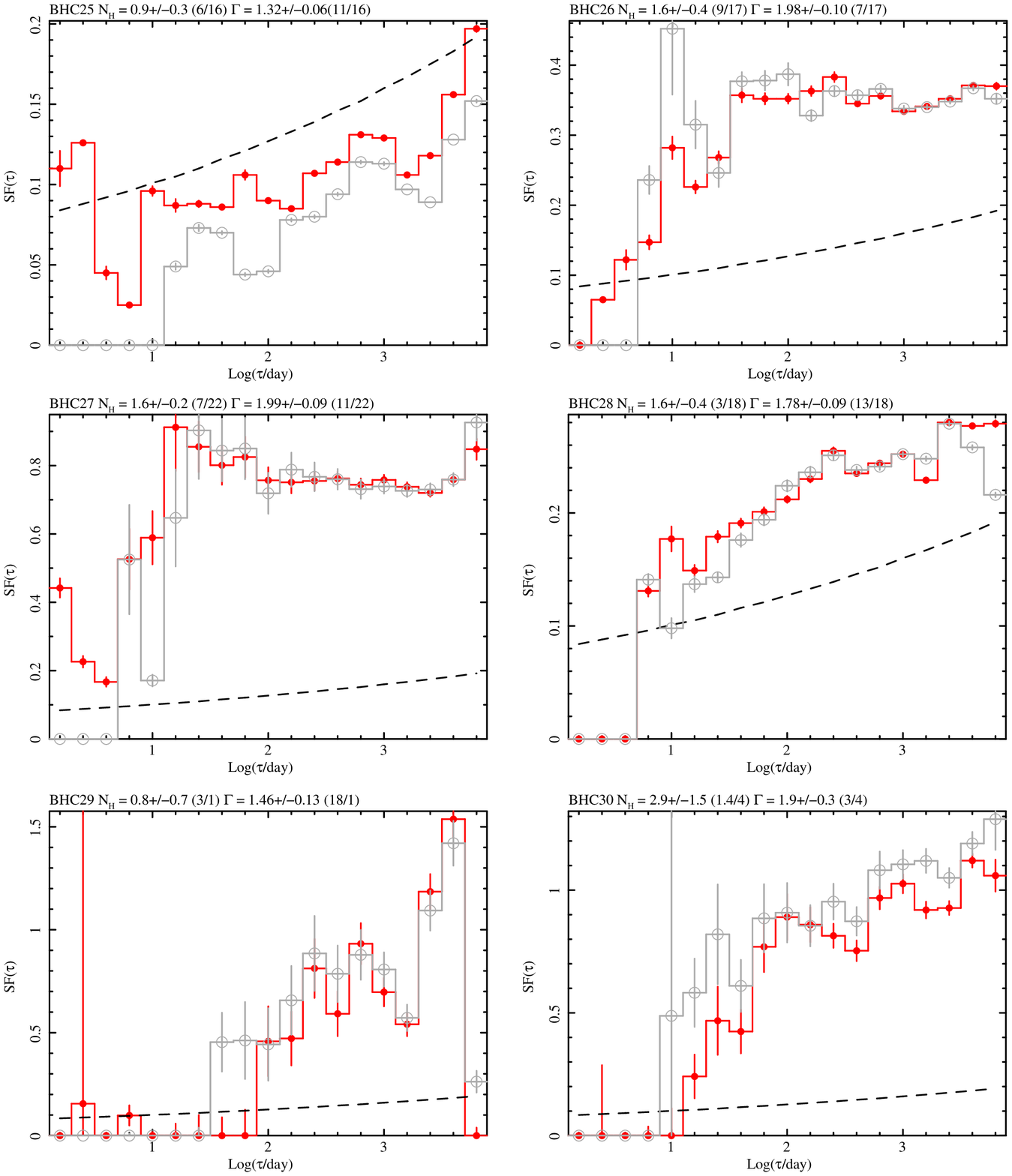}
\caption{continued}
\end{figure*}

\addtocounter{figure}{-1}
\begin{figure*}
\epsscale{1}
\plotone{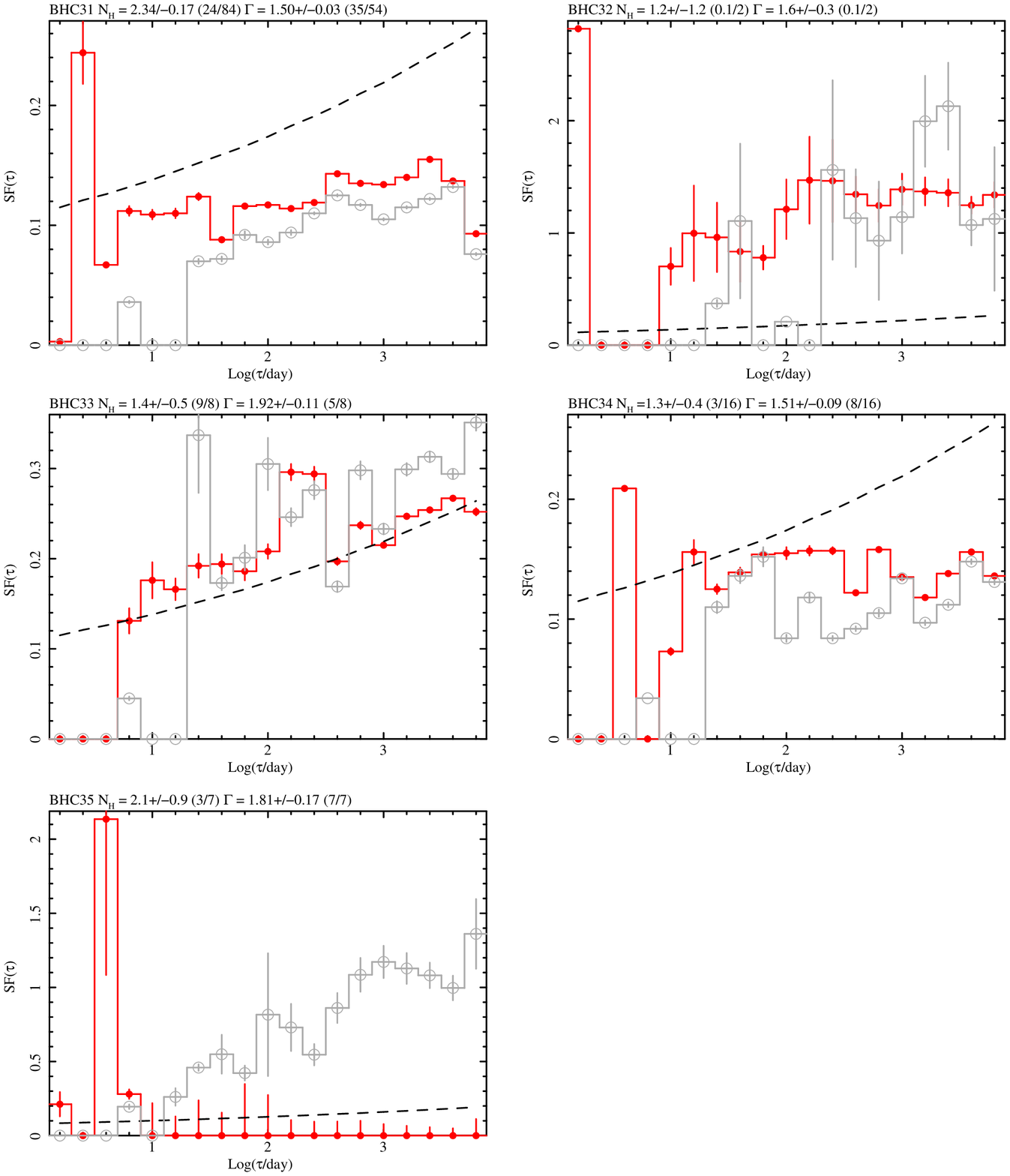}
\caption{continued}
\end{figure*}

\end{document}